\documentclass{aa}
\usepackage[varg]{txfonts}
\usepackage{graphicx}
\usepackage{natbib,twoopt}
\usepackage[breaklinks=true]{hyperref} 
\usepackage{lscape}
\bibpunct{(}{)}{;}{a}{}{,} 
\makeatletter
 \newcommandtwoopt{\citeads}[3][][]{\href{https://ui.adsabs.harvard.edu/abs/#3/abstract}%
 {\def\hyper@linkstart##1##2{}%
 \let\hyper@linkend\@empty\citealp[#1][#2]{#3}}}
 \newcommandtwoopt{\citepads}[3][][]{\href{https://ui.adsabs.harvard.edu/abs/#3/abstract}%
 {\def\hyper@linkstart##1##2{}%
 \let\hyper@linkend\@empty\citep[#1][#2]{#3}}}
 \newcommandtwoopt{\citetads}[3][][]{\href{https://ui.adsabs.harvard.edu/abs/#3/abstract}%
 {\def\hyper@linkstart##1##2{}%
 \let\hyper@linkend\@empty\citet[#1][#2]{#3}}}
 \newcommandtwoopt{\citeyearads}[3][][]%
 {\href{https://ui.adsabs.harvard.edu/abs/#3/abstract}
 {\def\hyper@linkstart##1##2{}%
 \let\hyper@linkend\@empty\citeyear[#1][#2]{#3}}}
\makeatother

\begin{document}

   \title{Centaur 2013~VZ$_{70}$: Debris from Saturn's irregular moon population?} 
   \author{C. de la Fuente Marcos$^{1}$
           \and
           R. de la Fuente Marcos$^{2}$}
   \authorrunning{C. de la Fuente Marcos \and R. de la Fuente Marcos}
   \titlerunning{Centaur 2013~VZ$_{70}$: Saturn's moon population debris?}
   \offprints{C. de la Fuente Marcos, \email{nbplanet@ucm.es}}
   \institute{$^{1}$ Universidad Complutense de Madrid,
              Ciudad Universitaria, E-28040 Madrid, Spain \\
              $^{2}$AEGORA Research Group,
              Facultad de Ciencias Matem\'aticas,
              Universidad Complutense de Madrid,
              Ciudad Universitaria, E-28040 Madrid, Spain}
   \date{Received 6 September 2021 / Accepted 8 October 2021}

   \abstract
      {Saturn has an excess of irregular moons. This is thought to be the 
       result of past collisional events. Debris produced during such episodes 
       in the neighborhood of a host planet can evolve into co-orbitals trapped 
       in quasi-satellite and/or horseshoe resonant states. A recently 
       announced centaur, 2013~VZ$_{70}$, follows an orbit that could be 
       compatible with those of prograde Saturn's co-orbitals.
       }
      {We perform an exploration of the short-term dynamical evolution of
       2013~VZ$_{70}$ to confirm or reject a co-orbital relationship with 
       Saturn. A possible connection with Saturn's irregular moon population is 
       also investigated.
       }
      {We studied the evolution of 2013~VZ$_{70}$ backward and forward in time 
       using $N$-body simulations, factoring uncertainties into the 
       calculations. We computed the distribution of mutual nodal distances 
       between this centaur and a sample of moons.  
       }
      {We confirm that 2013~VZ$_{70}$ is currently trapped in a horseshoe 
       resonant state with respect to Saturn but that it is a transient 
       co-orbital. We also find that 2013~VZ$_{70}$ may become a 
       quasi-satellite of Saturn in the future and that it may experience brief 
       periods of capture as a temporary irregular moon. This centaur might 
       also pass relatively close to known irregular moons of Saturn.
       }
      {Although an origin in trans-Neptunian space is possible, the hostile 
       resonant environment characteristic of Saturn's neighborhood favors a 
       scenario of in situ formation via impact, fragmentation, or tidal 
       disruption as 2013~VZ$_{70}$ can experience encounters with Saturn at 
       very low relative velocity. An analysis of its orbit within the context 
       of those of the moons of Saturn suggests that 2013~VZ$_{70}$ could be 
       related to the Inuit group, particularly Siarnaq, the largest and 
       fastest rotating member of the group. Also, the mutual nodal distances 
       of 2013~VZ$_{70}$ and the moons Fornjot and Thrymr are below the first 
       percentile of the distribution.
       }


   \maketitle

   \section{Introduction\label{Intro}}
      \citet{2021PSJ.....2..158A} found that Saturn has an excess of irregular moons compared to expectations based on Jupiter's irregular 
      moon population. They interpreted this excess as resulting from relatively recent collisional events. 

      Although not gravitationally bound to a host planet like a natural satellite, co-orbitals populate the 1:1 mean motion resonance that 
      makes them complete one revolution around a central star in almost exactly one sidereal orbital period of their host (see for example 
      \citealt{1999ssd..book.....M}). \citet{2018FrASS...5...13J} have argued that within the inner Solar System some co-orbital minor 
      bodies of Venus, Earth, and Mars may have an origin as impact ejecta. Such objects tend to approach their host planets at speeds close 
      to the planetary escape velocity. In general, catastrophic disruptions of small bodies in the neighborhood of a host planet may lead 
      to the production of co-orbitals and transient moons.

      Among the known planets of the Solar System and excluding Mercury, which has neither known natural satellites nor co-orbitals, Saturn 
      has the largest number of known moons (see for example \citealt{2021PSJ.....2..158A}) but also the smallest number of documented 
      co-orbitals (see for example \citealt{2006Icar..184...29G}). Saturn co-orbitals are indeed rare, and the dynamically hostile resonant 
      environment characteristic of Saturn's neighborhood may lead to their quick removal when they are either captured from the population 
      of Saturn-crossing minor bodies -- the centaurs and certain comets -- or produced in situ via impacts, fragmentations, or tidal 
      disruptions within Saturn's population of irregular moons. Although the origin of this population is still 
      disputed, \citet{2008MNRAS.391.1029T,2009MNRAS.392..455T} explored the evolution of the irregular satellites of Saturn, concluding 
      that the collisional capture scenario may explain their origin. 

      In this work we perform an exploration of the short-term dynamical evolution of 2013~VZ$_{70}$, a recently 
      announced\footnote{\url{https://minorplanetcenter.net/mpec/K21/K21Q55.html}} centaur that follows an orbit that could be compatible 
      with those of prograde Saturn's co-orbitals. This paper is organized as follows. In Sect.~\ref{Data} we provide the context of our 
      research, review our methodology, and present the data and tools used in our analyses. In Sect.~\ref{Results} we apply our 
      methodology and, in Sect.~\ref{Discussion}, discuss its results. Our conclusions are summarized in Sect.~\ref{Conclusions}.

   \section{Context, methods, and data\label{Data}}
      In the following, we provide some theoretical background to help navigate the reader through the presented results as well as the 
      basic details of our approach and the data and the tools used to obtain the results. 

      \subsection{Context}
         Co-orbital objects are engaged in a 1:1 mean-motion resonance with a host (see for example \citealt{1999ssd..book.....M}). In the 
         Solar System, co-orbital minor bodies go around the Sun in almost exactly one sidereal orbital period of a planetary host. There 
         are four main resonant states. Co-orbitals that follow prograde or direct paths (orbital inclination, $i$, $<90$\degr) can describe 
         tadpole (trojans), horseshoe, or quasi-satellite orbits; those in retrograde trajectories ($i>90$\degr) have trisectrix orbits,
         and this case is called the 1:$-$1 mean-motion resonance \citep{2017Natur.543..635M}. These orbital shapes are observed in a frame 
         of reference centered on the Sun and rotating with the host planet, projected onto the ecliptic plane. In the case of prograde 
         co-orbitals, hybrids of the three fundamental resonant states are possible \citep{1999PhRvL..83.2506N,2000CeMDA..76..131N}, as are 
         transitions between the various co-orbital states, elementary or hybrid \citep{1999Icar..137..293N,2000CeMDA..76..131N}. The 
         retrograde co-orbital problem has been further studied by, for example, \citet{2019MNRAS.490.3799M} and \cite{2020AJ....160..257S}.

         Co-orbital bodies are customarily identified by studying the evolution over time of a critical angle, whose value oscillates or 
         librates when the object under study is engaged in resonant behavior. For prograde orbits, this key parameter is the difference 
         between the mean longitude of the minor body (asteroid or comet) and that of its host planet. The mean longitude is given by 
         $\lambda=\Omega+\omega+M$, where $\Omega$ is the longitude of the ascending node, $\omega$ is the argument of perihelion, and $M$ 
         is the mean anomaly (see for example \citealt{1999ssd..book.....M}). For Saturn, the critical angle is 
         $\lambda_{\rm r}=\lambda-\lambda_{\rm S}$. When the value of $\lambda_{\rm r}$ oscillates about 0{\degr}, the body is a 
         quasi-satellite (or retrograde satellite, but it is not gravitationally bound) to the planet (see for example 
         \citealt{2006MNRAS.369...15M,2014CeMDA.120..131S}).\ When the libration is about 180{\degr}, often with an amplitude much wider than 
         180{\degr}, the minor body follows a horseshoe path. If it librates around 60{\degr}, the object is called an L$_4$ trojan and leads 
         the planet in its orbit.\ When it librates around $-$60{\degr} (or 300{\degr}), it is an L$_5$ trojan, and it trails the planet (see 
         for example \citealt{1999ssd..book.....M}). For the 1:$-$1 mean-motion resonance, the mean longitude of the retrograde body is 
         given by $\lambda^{*}=-\Omega+\omega+M$, and the critical angle is $\lambda_{\rm r}=\lambda^{*}-\lambda_{\rm S}-2\varpi^{*}$, where 
         $\varpi^{*}=\omega-\Omega$ and $\lambda_{\rm r}$ oscillates about 0{\degr} or 180{\degr} \citep{2013CeMDA.117..405M,
         2013MNRAS.436L..30M}. 

         The existence of Saturn's co-orbitals, particularly trojans, has been studied for decades (see for example 
         \citealt{1973AJ.....78..316E,1989AJ.....97..900I,1996Icar..121...88D,2000AJ....119.1978W, 2001MNRAS.322L..17M,2002ApJ...579..905M,
         2002Icar..160..271N,2014MNRAS.437.1420H,2019MNRAS.488.2543H}). Most studies concluded that, in general, the stability of Saturn's 
         co-orbitals is significantly weaker than that of their Jovian counterparts. \citet{2006Icar..184...29G} pointed out that 15504 
         (1999~RG$_{33}$) could be a quasi-satellite of Saturn, but \citet{2016MNRAS.462.3344D} used an improved orbit determination to show 
         that 15504 is a co-orbital but not a quasi-satellite. Centaur 63252 (2001~BL$_{41}$) is a true transient quasi-satellite of Saturn 
         \citep{2016MNRAS.462.3344D}. \citet{2018A&A...617A.114L} found several robust candidates to being trapped in the 1:$-$1 mean-motion 
         resonance with Saturn: centaurs 2006~RJ$_{2}$, 2006~BZ$_{8}$, 2017~SV$_{13}$, and 2012~YE$_{8}$.

      \subsection{Methodology}
         The assessment of the past, present, and future orbital evolution of 2013~VZ$_{70}$ should be based on the analysis of results from 
         a representative sample of $N$-body simulations that take the uncertainties in the orbit determination into account. Here, we 
         carried out such calculations using a direct $N$-body code implemented by \citet{2003gnbs.book.....A} that is publicly available 
         from the website of the Institute of Astronomy of the University of Cambridge.\footnote{\url{http://www.ast.cam.ac.uk/~sverre/web/pages/nbody.htm}} 
         This software uses the Hermite integration scheme devised by \citet{1991ApJ...369..200M}. Results from this code were extensively 
         discussed by \citet{2012MNRAS.427..728D}. Our calculations included the perturbations by the eight major planets, the Moon, the 
         barycenter of the Pluto-Charon system, and the three largest asteroids, (1) Ceres, (2) Pallas, and (4) Vesta. When studying the 
         dynamics of minor bodies in the region of the giant planets, it is particularly important to include all of them in the physical 
         model because they form a strongly coupled resonant subsystem \citep{1999Icar..139..336I,2002MNRAS.336..483I,2007PASJ...59..989T}. 
         Jupiter's and Saturn's semimajor axes experience a periodic modulation induced by a near 5:2 mean-motion resonance, a phenomenon
         known as the ``Great Inequality'' (see for example \citealt{1971NASTN6279.....M,2020AJ....160..232Z}). Results obtained under 
         three- or four-body approximations may not be applicable to objects such as 2013~VZ$_{70}$. 
         
      \subsection{Data, data sources, and tools}
         The discovery of 2013~VZ$_{70}$ was announced on 2021 August 23 \citep{2021MPEC....Q...55B}, but it had first been observed on 2013 
         November 1 by the Outer Solar System Origins Survey (OSSOS),\footnote{\url{http://www.ossos-survey.org/}} which is a survey 
         aimed at carefully sampling important trans-Neptunian populations in order to verify various models of giant planet migration 
         during the early stages of the formation of the Solar System \citep{2016AJ....152...70B}. Although not formally announced until 
         2021, this object had previously been mentioned by \citet{2018DPS....5030509A,2020DPS....5220606A}. Its current orbit determination 
         is shown in Table~\ref{elements}, and it is based on 36 observations spanning a data arc of 946~d. Its trajectory is unusual among 
         those of objects in the orbital neighborhood of Saturn (but not gravitationally bound to it) as it has both low eccentricity, $e$, 
         and inclination, $i$. Saturn's co-orbital zone goes, in terms of semimajor axis, $a$, from $\sim$9~AU to $\sim$10~AU (see for 
         example \citealt{2014MNRAS.437.1420H}). Therefore, considering the values of $a$ and $e$ in Table~\ref{elements}, 2013~VZ$_{70}$ 
         is a robust candidate to being engaged in a 1:1 mean-motion resonance with Saturn.
%
%
      \begin{table}
         \centering
         \fontsize{8}{11pt}\selectfont
         \tabcolsep 0.15truecm
         \caption{\label{elements}Heliocentric Keplerian orbital elements of 2013~VZ$_{70}$.
                 }
         \begin{tabular}{ccc}
            \hline\hline
             Parameter                                         &   &   Value$\pm\sigma$    \\
            \hline
             Semimajor axis, $a$ (AU)                         & = &   9.1457$\pm$0.0003   \\
             Eccentricity, $e$                                 & = &   0.09448$\pm$0.00002 \\
             Inclination, $i$ (\degr)                          & = &  12.05274$\pm$0.00009 \\
             Longitude of the ascending node, $\Omega$ (\degr) & = & 215.17744$\pm$0.00012 \\
             Argument of perihelion, $\omega$ (\degr)          & = & 245.302$\pm$0.014     \\
             Mean anomaly, $M$ (\degr)                         & = &  34.155$\pm$0.009     \\
             Perihelion, $q$ (AU)                              & = &   8.2816$\pm$0.0002   \\
             Aphelion, $Q$ (AU)                                & = &  10.0098$\pm$0.0003   \\
             Absolute magnitude, $H$ (mag)                     & = &  13.7$\pm$0.3         \\
            \hline
         \end{tabular}
         \tablefoot{Values are displayed together with the 1$\sigma$ uncertainty and are referred to epoch JD 2459396.5, which corresponds 
                    to 0:00 on 2021 July 1 Barycentric Dynamical Time (TDB) (J2000.0 ecliptic and equinox). Source: JPL's SBDB (solution 
                    date, 2021 August 24 07:42:02 PDT).
                   }
      \end{table}
%
%

         Here, we work with publicly available data (orbit determinations, input Cartesian vectors, and ephemerides) from the Jet Propulsion 
         Laboratory (JPL) Small-Body Database (SBDB)\footnote{\url{https://ssd.jpl.nasa.gov/tools/sbdb_lookup.html\#/}} and the {\tt Horizons} 
         online Solar System data and ephemeris computation service,\footnote{\url{https://ssd.jpl.nasa.gov/horizons/}} both provided by 
         the Solar System Dynamics Group\footnote{\url{https://ssd.jpl.nasa.gov/}} \citep{2011jsrs.conf...87G,2015IAUGA..2256293G}. Most 
         data were retrieved from JPL's SBDB and {\tt Horizons} using tools provided by the Python package Astroquery 
         \citep{2019AJ....157...98G}. 

         In order to evaluate data clustering for the sample of Saturnian moons in Sect.~4.1, we applied the unsupervised machine-learning 
         algorithm $k$-means++ as implemented by the Python library Scikit-learn \citep{2012arXiv1201.0490P}; we used the elbow method to 
         determine the optimal value of clusters, $k$ (for details, see \citealt{2021MNRAS.501.6007D}). Some figures have been produced using 
         the Matplotlib library \citep{2007CSE.....9...90H} and statistical tools provided by NumPy \citep{2011CSE....13b..22V,
         2020Natur.585..357H}. Sets of bins in histograms were computed using NumPy via the application of the Freedman and Diaconis rule 
         \citep{Freedman1981}; instead of using frequency-based histograms, we consider counts to form a probability density such that the 
         area under the histogram will sum to one.

   \section{Results\label{Results}} 
\begin{figure}
         \centering
         \includegraphics[width=\linewidth]{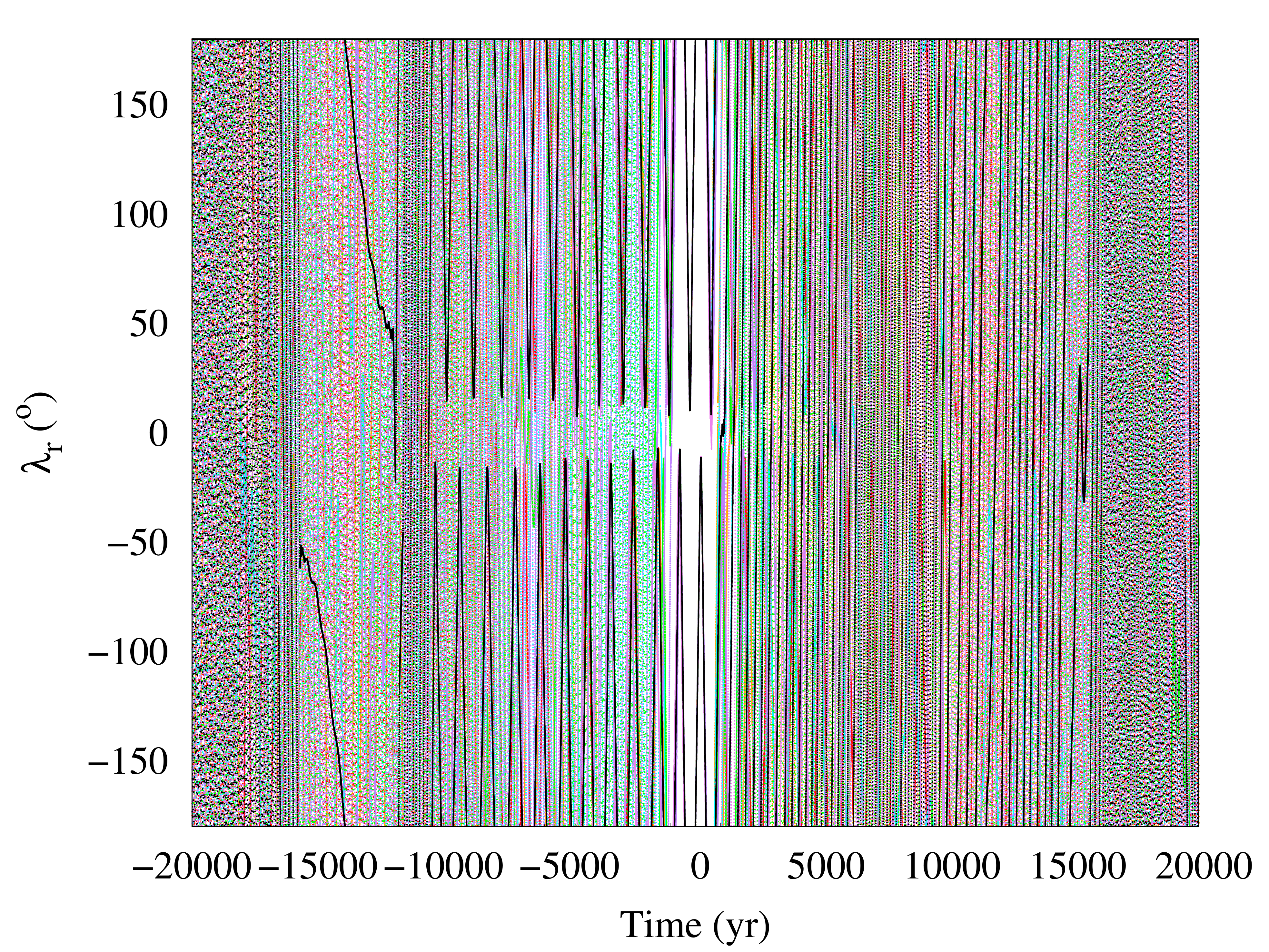}
         \includegraphics[width=\linewidth]{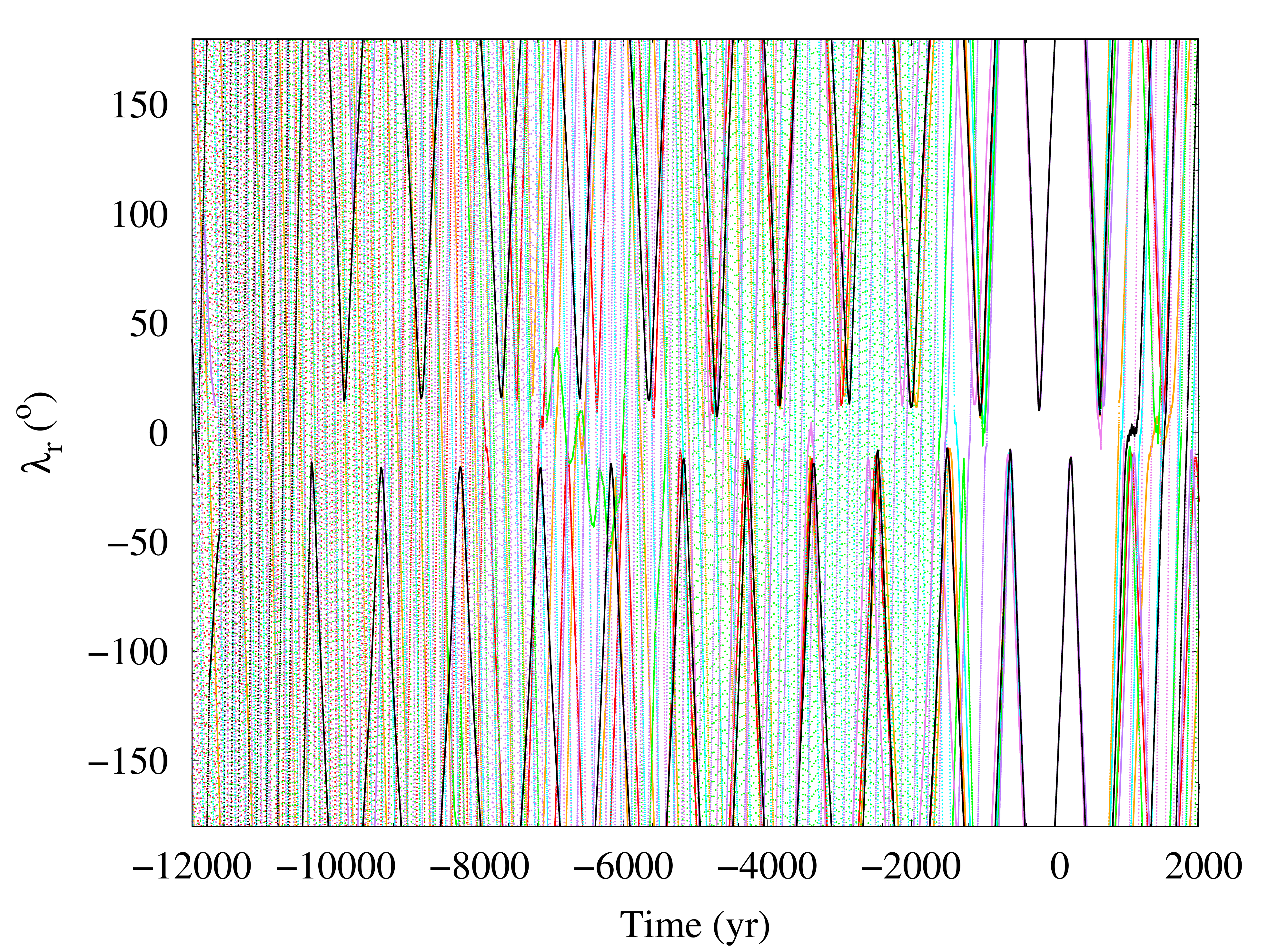}
         \includegraphics[width=\linewidth]{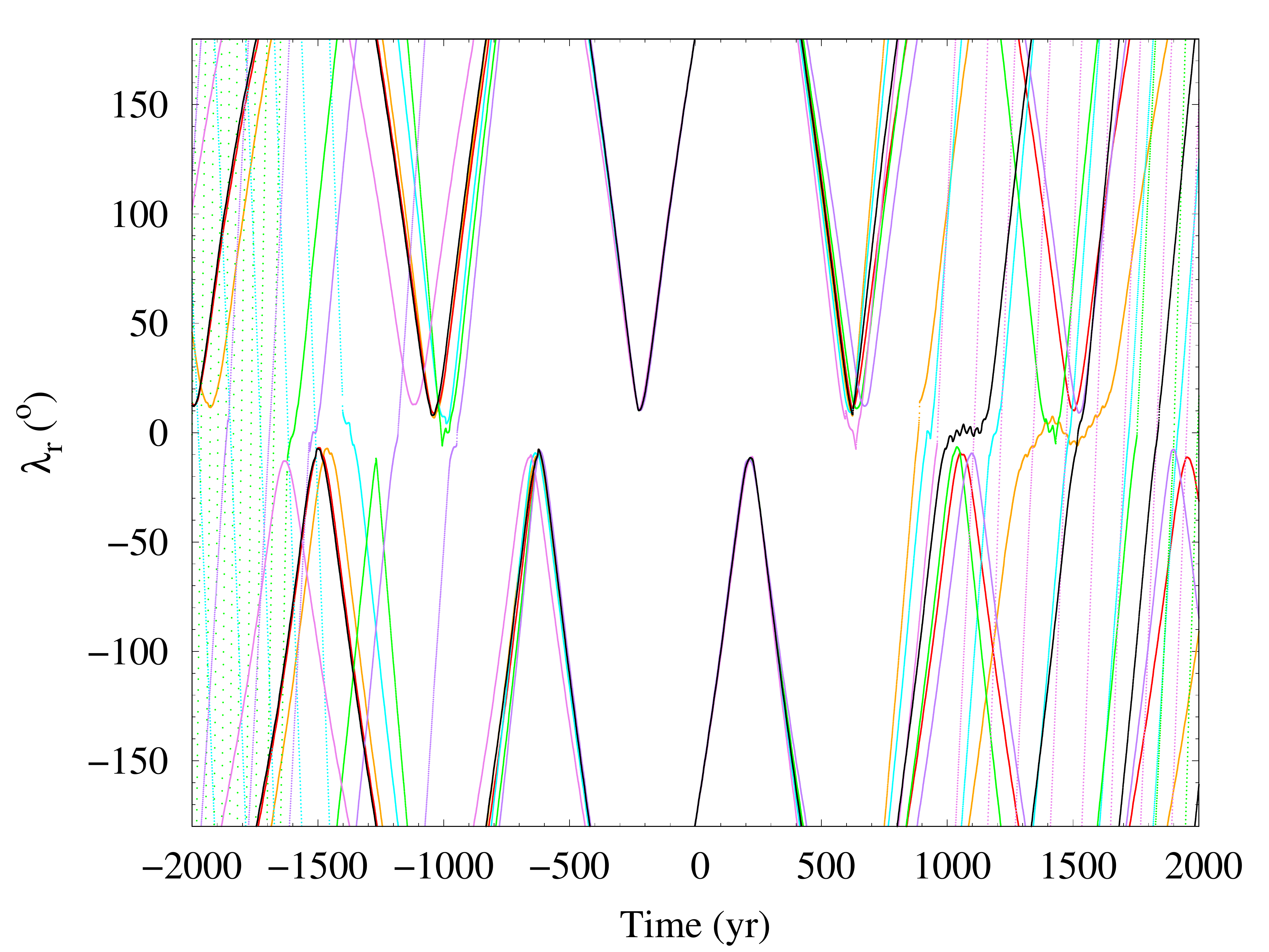}
         \caption{Evolution of the value of the relative mean longitude, $\lambda_{\rm r}$, of 2013~VZ$_{70}$ with respect to Saturn. The 
                  results corresponding to the nominal orbit are shown in black, and the results of control orbits with Cartesian vectors 
                  separated by $+$1$\sigma$ (in red), $-$1$\sigma$ (in orange), $+$3$\sigma$ (in green), $-$3$\sigma$ (in cyan), 
                  $+$9$\sigma$ (in purple), and $-$9$\sigma$ (in violet) from the nominal values are also displayed. The output time-step 
                  size is 0.5~yr. The three panels show different time ranges to make the visualization easier. The input data (Cartesian 
                  vectors for all the Solar System objects) have JPL's SBDB as a source and are referred to epoch 2459396.5 Barycentric 
                  Dynamical Time, which is also the origin of time in the calculations.
                 }
         \label{evolution}
      \end{figure}

      Figure~\ref{evolution} shows the results of our integrations, focusing on the evolution of $\lambda_{\rm r}$ and summarizing our 
      findings for representative orbits, the nominal one (in black) and those with Cartesian vectors separated $\pm$1$\sigma$, 
      $\pm$3$\sigma$, and $\pm$9$\sigma$ from the nominal values (see Table~\ref{vector2013VZ70} for input values). The top panel shows that 
      the orbital evolution of 2013~VZ$_{70}$ was as chaotic in the past as it will be in the future. The middle panel shows that the 
      nominal orbit in Table~\ref{elements} displays horseshoe behavior for about 1.2$\times$10$^{4}$~yr, but most control orbits only 
      remained in the horseshoe resonant state for a fraction of that time. It also shows that the evolution of the nominal orbit of 
      2013~VZ$_{70}$ was far more stable in the past than it will be in the future. 
%
%
%
%

      However, the bottom panel provides the most comprehensive evaluation of the current dynamical status of this object. All the control 
      orbits, even those separated by $\pm$9$\sigma$ from the nominal values, show consistent co-orbital behavior in the time interval 
      ($-$1000, 1000)~yr. Any prediction beyond that time window based on the current orbit determination is very uncertain. All the control 
      orbits show that 2013~VZ$_{70}$ remains trapped in the horseshoe resonant state during the interval ($-$1000,~900)~yr. Therefore, we 
      can confirm that 2013~VZ$_{70}$ is indeed a temporary or transient co-orbital to Saturn and that it currently follows a horseshoe-type 
      path in a frame of reference centered on the Sun and rotating with Saturn. This centaur can experience deep and slow encounters with 
      Saturn and become temporarily captured by the ringed planet as a short-lived moon (see below). For the nominal orbit, the bottom panel 
      of Fig.~\ref{evolution} shows that after one such flyby, 2013~VZ$_{70}$ becomes a quasi-satellite of Saturn for nearly 200~yr, 
      starting at about 975~yr into the future. This episode is analyzed in more detail below. 

\begin{figure}
        \centering
        \includegraphics[width=\linewidth]{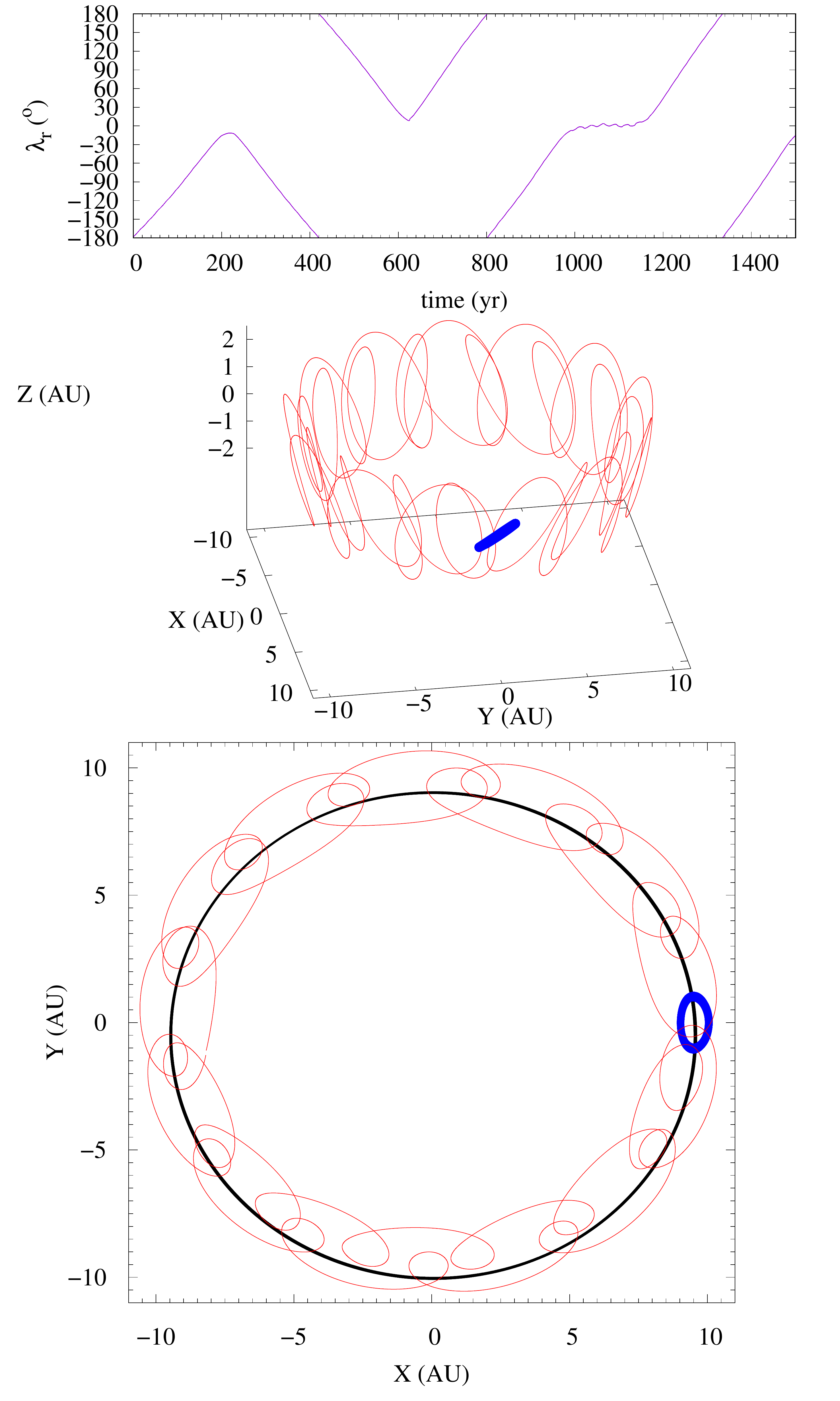}
        \caption{Future horseshoe behavior for the nominal orbit. Top panel: Future evolution of the value of the relative mean longitude, 
                 $\lambda_{\rm r}$, with respect to Saturn for the nominal orbit of 2013~VZ$_{70}$. Middle panel: Trajectory in 
                 three-dimensional space during the time interval (0, 800)~yr in a frame of reference centered on the Sun and rotating with 
                 Saturn. Bottom panel: The path followed by 2013~VZ$_{70}$ (which moves counterclockwise) is in a frame of reference 
                 centered on the Sun and rotating with Saturn, projected onto the ecliptic plane, during the same time interval. Since 
                 Saturn follows an eccentric orbit, it is represented in the panels by a blue trace (middle) or ellipse (bottom). The output 
                 time-step size is 0.01~yr.   
                }
        \label{futureevolution}
     \end{figure}

      Figure~\ref{futureevolution} illustrates how 2013~VZ$_{70}$ will pursue its horseshoe path as Saturn's co-orbital in the future for 
      the nominal orbit, and for other control orbits the evolution is very similar. The horseshoe orbital period is nearly 900~yr, which is 
      also the period of the perturbation associated with the Great Inequality (see for example \citealt{1971NASTN6279.....M}). After a 
      close encounter with Saturn and for the nominal orbit, 2013~VZ$_{70}$ will transition from the horseshoe resonant state to a 
      quasi-satellite orbit that may approach Saturn more closely and more slowly.
%
%
%
%

\begin{figure}
        \centering
        \includegraphics[width=\linewidth]{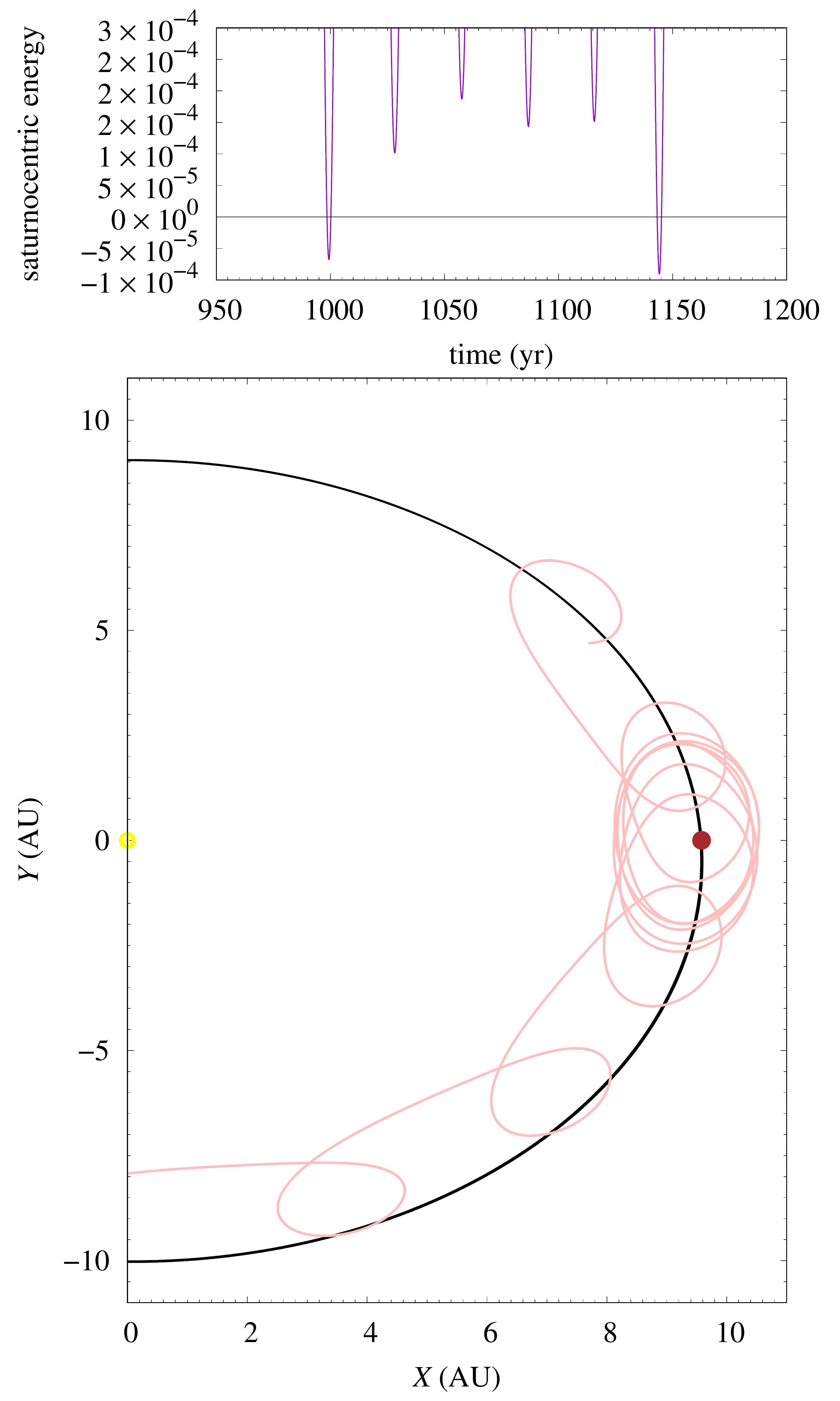}
        \caption{Top panel: Evolution of the Keplerian Saturnocentric energy of 2013~VZ$_{70}$. Satellite captures happen when the relative 
                 binding energy becomes negative. The unit of energy is such that the unit of mass is 1~$M_{\odot}$, the unit of distance is 
                 1~AU, and the unit of time is one sidereal year divided by 2$\pi$. Bottom panel: Path followed by 2013~VZ$_{70}$ (which 
                 moves counterclockwise) in a frame of reference centered on the Sun (in yellow) and rotating with Saturn (in brown, its 
                 orbit in black), projected onto the ecliptic plane, during the time interval (950, 1200)~yr, shown in pink. The output 
                 time-step size is 0.01~yr.
                }
        \label{energy}
     \end{figure}

      The top panel of Figure~\ref{energy} shows that the Keplerian Saturnocentric energy of 2013~VZ$_{70}$ (relative binding energy) became 
      negative during the abovementioned encounter (twice, for about 3~yr each), so it became a temporary (retrograde) irregular moon of 
      Saturn. However, the relative binding energy was not negative for the full length of a loop around Saturn (the loops are traveled in 
      the clockwise direction), so, following the terminology in \citet{2017Icar..285...83F}, we may speak of a temporarily captured flyby as 
      the centaur does not complete an entire loop around Saturn with negative relative binding energy. Other control orbits (not shown) may 
      lead to temporary captures (temporarily captured flybys and orbiters) when integrating both into the past and forward in time 
      (including recurrent episodes).
%
%
%
%

   \section{Discussion\label{Discussion}} 
      \citet{2018DPS....5030509A,2020DPS....5220606A,2021PSJ.....2..212A} favor an origin for 2013~VZ$_{70}$ in the trans-Neptunian 
      populations. One of Saturn's irregular moons, Phoebe, is believed to be a captured centaur with a trans-Neptunian origin (see for 
      example \citealt{2005Natur.435...69J,2007ARA&A..45..261J}). Although such an origin is indeed possible, here we argue that the hostile 
      resonant environment characteristic of Saturn's neighborhood favors a scenario of in situ formation via impact, fragmentation, or 
      tidal disruption within the population of irregular moons as 2013~VZ$_{70}$ can experience encounters with Saturn at very low relative 
      velocity (see above). Supporting arguments for a scenario of in situ formation come from two sides. In both cases, we used a relevant 
      sample of natural satellites of Saturn. Given the fact that the orbit determinations of some moons of Saturn are somewhat poor and 
      that their uncertainties are far from well characterized, we initially restricted our statistical analyses to the nominal orbits. 

      \subsection{Natural satellites of Saturn: Orbital context}
\begin{figure}
        \centering
        \includegraphics[width=\linewidth]{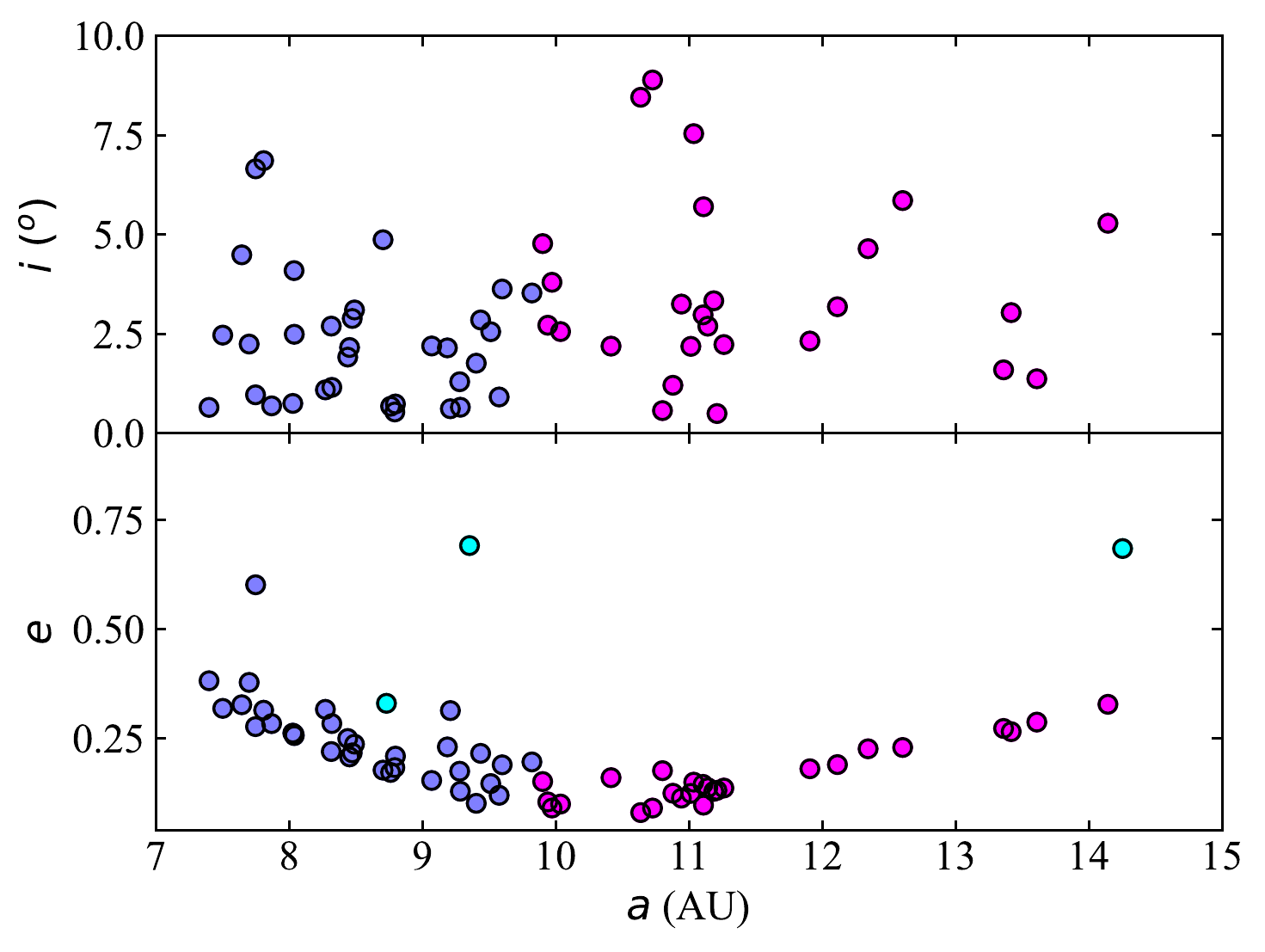}
        \caption{Color-coded clusters generated by the $k$-means++ algorithm applied to the data set made of 64 Saturnian moons. The figure
                 shows the heliocentric values ($a$, $e$, $i$). The range in $i$ has been restricted to the one relevant to this work. 
                 Heliocentric Keplerian orbital elements refer to epoch JD 2459396.5. Source: JPL's SBDB and {\tt Horizons}.
                }
        \label{moons}
     \end{figure}

         We retrieved the heliocentric orbital elements of the known 
         natural satellites of Saturn from JPL's SBDB for the epoch JD 2459396.5. Our sample included 64 moons with heliocentric $e<1$ and 
         $a<20$~AU. The subsample with $e<0.25$ is shown in Table~\ref{moonelements}. Figure~\ref{moons} shows the result of applying the 
         $k$-means++ algorithm and the elbow method to the data set of Saturn's moon. Three clusters are found: The plum and fuchsia points 
         include members of the Inuit and Gallic groups that follow Saturnocentric prograde orbits as well as members of the Norse group 
         that follow Saturnocentric retrograde orbits (see Table~\ref{moonelements}), and the azure points include inner moons such as 
         Atlas, Pan, and Polydeuces, but also Rhea (not shown in the top panel).

%
%
      \begin{table}
         \centering
         \fontsize{8}{11pt}\selectfont
         \tabcolsep 0.15truecm
         \caption{\label{moonelements}Heliocentric Keplerian orbital elements of Saturnian moons with $e<0.25$.
                 }
         \begin{tabular}{ccccccc}
            \hline\hline
             Moon        & Group   & $a$     & $e$    & $i$     & $\Omega$ & $\omega$ \\
                         &         & (AU)    &        & (\degr) & (\degr)  & (\degr)  \\
            \hline
             Tarqeq      & Inuit   & 10.6373 & 0.0798 & 8.4581  & 124.6886 & 135.9928 \\
             S/2004 S 31 & Inuit   & 10.7263 & 0.0902 & 8.8928  & 125.3434 & 139.3948 \\
             S/2004 S 13 & Norse   &  9.9718 & 0.0903 & 3.8017  & 119.3540 &  97.1457 \\
             Siarnaq     & Inuit   & 11.1085 & 0.0968 & 5.7029  & 116.5324 & 178.2054 \\
             S/2004 S 22 & Norse   & 10.0361 & 0.0994 & 2.5610  & 114.0951 & 104.8162 \\
             S/2004 S 12 & Norse   &  9.4036 & 0.1010 & 1.7650  &  98.7427 &  79.3981 \\
             S/2004 S 28 & Norse   &  9.9404 & 0.1043 & 2.7207  & 118.9490 &  96.0909 \\
             S/2004 S 34 & Norse   & 10.9428 & 0.1130 & 3.2529  & 114.4503 & 170.0351 \\
             S/2006 S  1 & Norse   &  9.5755 & 0.1194 & 0.9149  &  19.8416 & 172.6876 \\ 
             S/2004 S 26 & Norse   & 11.0136 & 0.1238 & 2.1887  & 107.8656 & 166.0223 \\
             S/2004 S 17 & Norse   & 10.8789 & 0.1241 & 1.2094  &  91.8447 & 168.7292 \\
             Surtur      & Norse   & 11.1856 & 0.1285 & 3.3369  & 119.4188 & 197.0921 \\
             Hyrrokkin   & Norse   &  9.2844 & 0.1287 & 0.6561  & 343.7056 & 192.7894 \\ 
             S/2004 S 21 & Norse   & 11.2098 & 0.1307 & 0.4981  &  41.1908 & 278.4422 \\
             Paaliaq     & Inuit   & 11.2614 & 0.1360 & 2.2346  & 318.0790 &  32.7217 \\
             S/2004 S 25 & Norse   & 11.1409 & 0.1367 & 2.6967  & 116.3389 & 220.7748 \\
             Thrymr      & Norse   & 11.1055 & 0.1444 & 2.9811  & 115.8016 & 143.8882 \\          
             Kari        & Norse   &  9.5122 & 0.1463 & 2.5574  & 106.2945 &  84.4690 \\
             Albiorix    & Gallic  & 11.0350 & 0.1495 & 7.5428  & 123.4528 & 125.6591 \\              
             Ijiraq      & Inuit   &  9.9018 & 0.1509 & 4.7754  & 116.2447 & 294.6973 \\
             S/2004 S 35 & Norse   &  9.0692 & 0.1534 & 2.1981  & 111.6498 &  58.1992 \\
             S/2004 S 23 & Norse   & 10.4147 & 0.1598 & 2.1943  & 110.6376 & 281.0498 \\ 
             Jarnsaxa    & Norse   &  8.7613 & 0.1719 & 0.6805  &  78.0443 &  78.3403 \\
             S/2004 S 24 & Gallic  &  9.2794 & 0.1745 & 1.2985  & 123.0952 & 309.4057 \\
             Bergelmir   & Norse   & 10.8008 & 0.1759 & 0.5717  &  16.2162 &   3.7236 \\ 
             S/2004 S 38 & Norse   &  8.7042 & 0.1770 & 4.8721  & 118.8771 & 354.5464 \\
             Greip       & Norse   & 11.9067 & 0.1803 & 2.3226  & 110.5977 & 175.1823 \\ 
             S/2004 S 36 & Norse   &  8.7927 & 0.1829 & 0.5376  &  69.5592 &  20.9945 \\ 
             Skathi      & Norse   &  9.5980 & 0.1893 & 3.6319  & 115.1900 &  81.8631 \\
             S/2004 S 7  & Norse   & 12.1132 & 0.1898 & 3.1864  & 112.5916 & 202.4078 \\
             S/2004 S 27 & Norse   &  9.8213 & 0.1959 & 3.5323  & 114.9983 & 298.1294 \\
             Fornjot     & Norse   &  8.4532 & 0.2074 & 2.1604  & 106.6699 &  43.9892 \\
             Fenrir      & Norse   &  8.7972 & 0.2097 & 0.7378  &  65.1595 &  24.3685 \\
             S/2004 S 30 & Norse   &  8.4732 & 0.2168 & 2.8906  & 110.6679 &  45.4170 \\
             Ymir        & Norse   &  8.3159 & 0.2197 & 2.6986  & 112.2017 &  13.9596 \\
             S/2004 S 32 & Norse   & 12.3433 & 0.2260 & 4.6474  & 121.0811 & 219.1888 \\
             Erriapus    & Gallic  & 12.6026 & 0.2290 & 5.8587  & 120.9251 & 224.0407 \\
             S/2004 S 29 & Inuit   &  9.1870 & 0.2303 & 2.1507  & 100.5478 & 332.4760 \\
             S/2007 S 2  & Norse   &  8.4902 & 0.2365 & 3.1050  & 115.4799 & 340.9548 \\
             S/2004 S 39 & Norse   &  8.4395 & 0.2496 & 1.9188  & 103.1231 & 353.2404 \\
            \hline
         \end{tabular}
         \tablefoot{The orbits, sorted by $e$, are referred to epoch JD 2459396.5, which corresponds to 0:00 on 2021 July 1 TDB  (J2000.0 
                    ecliptic and equinox). Most moons with provisional designations may correspond to lost objects that may have to be
                    rediscovered (see for example \citealt{2012AJ....144..132J}). Source: JPL's SBDB and {\tt Horizons}.
                   }
      \end{table}
%
%
%
%
%
%
         Table~\ref{moonelements} shows that the heliocentric orbital elements of multiple irregular moons of Saturn resemble those of 
         2013~VZ$_{70}$ in Table~\ref{elements}. However, most moons with provisional designations may correspond to lost objects that may 
         have to be rediscovered (see for example \citealt{2012AJ....144..132J}). Therefore, and among the moons in Table~\ref{moonelements},
         we would like to single out Tarqeq and Siarnaq. Tarqeq (originally named S/2007 S 1 or Saturn LII; \citealt{2007IAUC.8836....1S,
         2007MPEC....G...38S}) has a size of about 6~km \citep{2019Icar..322...80D} and is a member of the Inuit group of irregular 
         satellites of Saturn that also follow a low-eccentricity, low-inclination heliocentric orbit (see Table~\ref{moonelements}), like 
         2013~VZ$_{70}$. Another object with similar orbital properties is Siarnaq (originally named S/2000 S 3 or Saturn XXIX; 
         \citealt{2000IAUC.7513....1G,2001MPEC....U...42H}), which is the largest member of the Inuit group of Saturnian satellites, at 
         about 39~km \citep{2015ApJ...809....3G}, and one of the fastest rotators, with a period of 10~h \citep{2019Icar..322...80D}. 

         Centaur 2013~VZ$_{70}$ could be similar in size to Tarqeq and other irregular satellites of Saturn, and its heliocentric orbit is 
         consistent in terms of $e$ and $i$ with those of other moons. We interpret these facts as supportive of an origin among one of the 
         groups of irregular satellites of Saturn.

      \subsection{Saturnian moons versus 2013~VZ$_{70}$: Relative nodal distances}
\begin{figure}
        \centering
        \includegraphics[width=\linewidth]{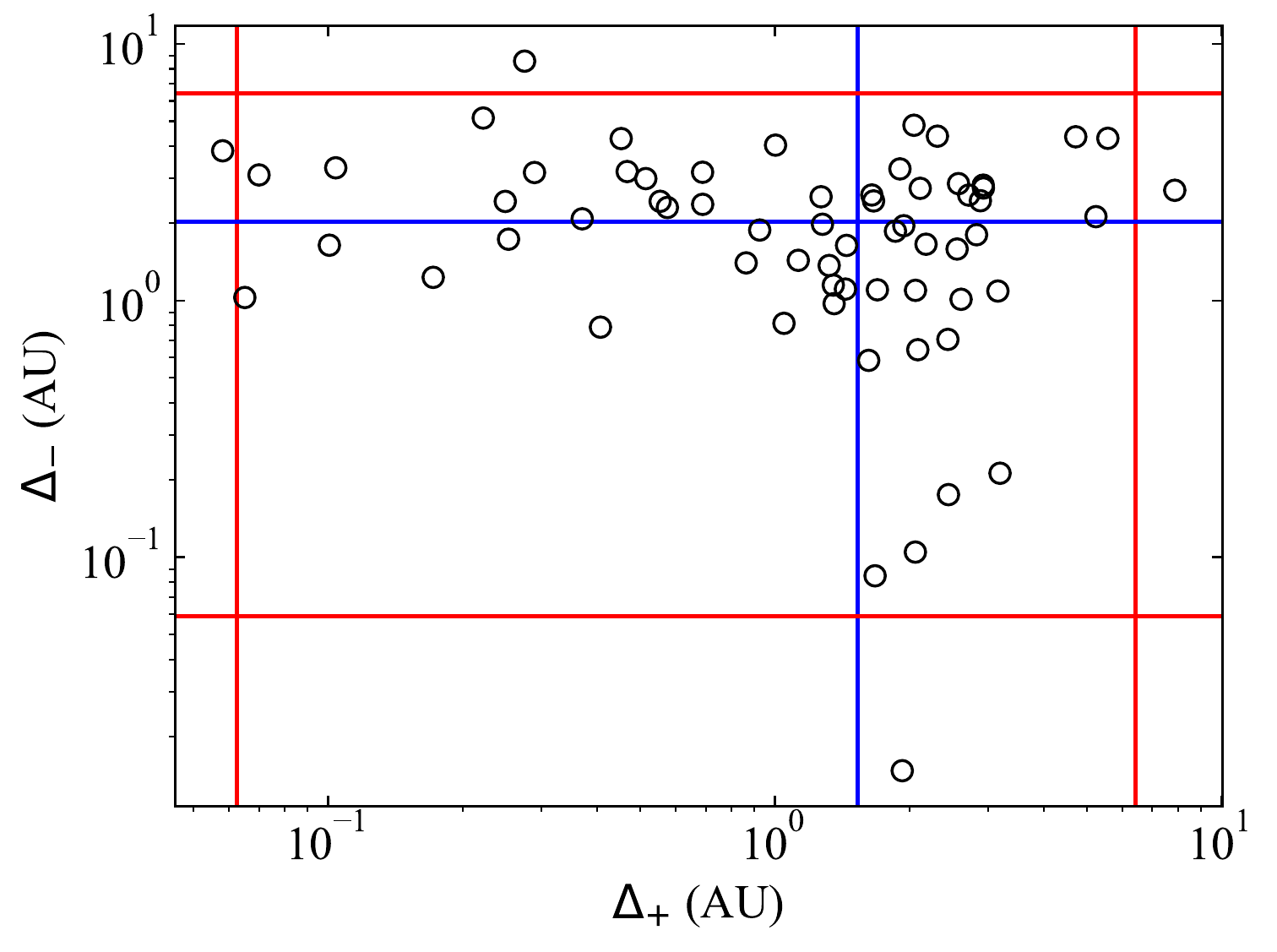}
        \caption{Distribution of mutual nodal distances (${\Delta}_{+}$, ascending; ${\Delta}_{-}$, descending) between 2013~VZ$_{70}$ and a 
                 sample of Saturn satellites. The median values are shown in blue and the 1st and 99th percentiles in red.
                }
        \label{moonnodes}
     \end{figure}

         We computed the distribution of the absolute values of the mutual nodal distances of 2013~VZ$_{70}$ and the sample of 
         Saturnian moons using Eqs.~16 and 17 from \citet{2017CeMDA.129..329S} -- ${\Delta}_{+}$ for the ascending nodes and ${\Delta}_{-}$ 
         for the descending nodes -- and data from JPL's SBDB and {\tt Horizons} referred to epoch JD 2459396.5. Our results are shown in 
         Fig.~\ref{moonnodes}, and they indicate that the heliocentric orbits of some irregular satellites of Saturn pass rather close to 
         the trajectory of 2013~VZ$_{70}$. The first percentile of the distribution of ${\Delta}_{+}$ is 0.062~AU, and the mutual nodal 
         distance between the ascending nodes of 2013~VZ$_{70}$ and Thrymr is 0.058~AU. On the other hand, the first percentile of the 
         distribution of ${\Delta}_{-}$ is 0.059~AU, and the mutual nodal distance between the descending nodes of 2013~VZ$_{70}$ and 
         Fornjot is 0.015~AU. Both cases are clear outliers. 
%
%
%
%

         Thrymr and Fornjot are members of the Norse group (see Table~\ref{moonelements}) of irregular satellites of Saturn that follow 
         Saturnocentric retrograde orbits. Thrymr (originally named S/2000 S 7 or Saturn XXX; \citealt{2001MPEC....X...20P}) is believed to 
         be debris released during an impact on Phoebe \citep{2019Icar..322...80D}. Fornjot (originally named S/2004 S 8 or Saturn XLII; 
         \citealt{2006MPEC....C...74J}) is one of the outermost natural satellites of Saturn and also one of the fastest rotators, with a 
         period of 7 or 9.5~h \citep{2019Icar..322...80D}. 

         We interpret the existence of these low probability, short mutual nodal distances as an indication of a collisional origin for 
         2013~VZ$_{70}$. However, for a given pair of objects, a present-day short mutual nodal distance does not imply an equally short 
         value in the past or the future. In addition, although it is true that close flybys take place in the vicinity of the mutual 
         nodes, short mutual nodal distances may not translate into actual flybys if there are active protection mechanisms such as 
         mean-motion or secular resonances like in the cases of Neptune's trojans or Pluto (see for example \citealt{1989Icar...82..200M,
         2001AJ....121.1155W}).  

      \subsection{Possible past encounters of 2013~VZ$_{70}$ with Fornjot and Thrymr}
         The present-day short mutual nodal distances of 2013~VZ$_{70}$ and Fornjot and Thrymr are indicative of a possible past collisional
         evolution of 2013~VZ$_{70}$. The assessment of the concurrent past orbital evolution of 2013~VZ$_{70}$ and Fornjot and Thrymr 
         should be based on the statistical analysis of results from a representative sample of $N$-body simulations. Unfortunately, in this
         case the effect of the uncertainties is not easy to include in the calculations. 

         The current orbit determination of 2013~VZ$_{70}$ in Table~\ref{elements} is based on data collected for less than 10\% of its 
         sidereal orbital period. Although it can certainly be improved, the relative errors in the orbital elements are in the range 
         10$^{-4}$--10$^{-5}$, and this makes any short-term ephemerides (for example, present-day Cartesian state vectors) computed from it
         reasonably robust. However, this centaur moves in a very unstable region, and the results presented in Sect.~\ref{Results} indicate 
         that control orbits relatively close to the nominal one lead to a rather different orbital evolution both into the past and forward 
         in time outside the time interval ($-$1000,~1000)~yr. In sharp contrast, the orbit determinations of known irregular moons of 
         Saturn have uncertainties affected by poorly characterized systematics present in sparse observation sets. Figure~1 in 
         \citet{2012AJ....144..132J} shows that the on-sky position uncertainty of these irregular moons may oscillate over time. Table~3 in 
         \citet{2012AJ....144..132J} shows that this uncertainty could be as high as 0{\farcs}7 and 5{\farcs}7 for Thrymr and Fornjot, 
         respectively, after three orbital periods beyond 2012 January. In such cases, there is no reliable procedure for computing meaningful 
         1$\sigma$ uncertainties for a given set of barycentric Cartesian state vectors. 

         The planetary satellite ephemeris SAT368 computed by R.A. Jacobson in 2014 updated the irregular satellites of Saturn with 
         Earth-based data through early 2014 and all Cassini imaging through 2013, and they are utilized by {\tt Horizons} to provide Cartesian 
         state vectors that can be used as input data to perform the required $N$-body simulations. {\tt Horizons} cannot provide estimates 
         of the associated 1$\sigma$ uncertainties, but the barycentric Cartesian state vectors are shown in Tables~\ref{vectorFornjot} and 
         \ref{vectorThrymr}. We used a sample of 1\,500~pairs of Gaussian-distributed control orbits (see Appendix~\ref{Adata}) based on 
         the Cartesian vectors in Tables~\ref{vector2013VZ70} and \ref{vectorFornjot} and integrated backwards in time for 1\,500~yr 
         (assuming uncertainties of 10\% and 5\% for Fornjot), considering the same physical model used in Sect.~\ref{Results} (in this 
         simplified set of experiments, the contribution of massive moons was neglected), and studied the distribution of minimum approach 
         distances (see \citealt{2021A&A...649A..85D} for additional details on this approach). 

\begin{figure*}
        \centering
        \includegraphics[width=0.49\linewidth]{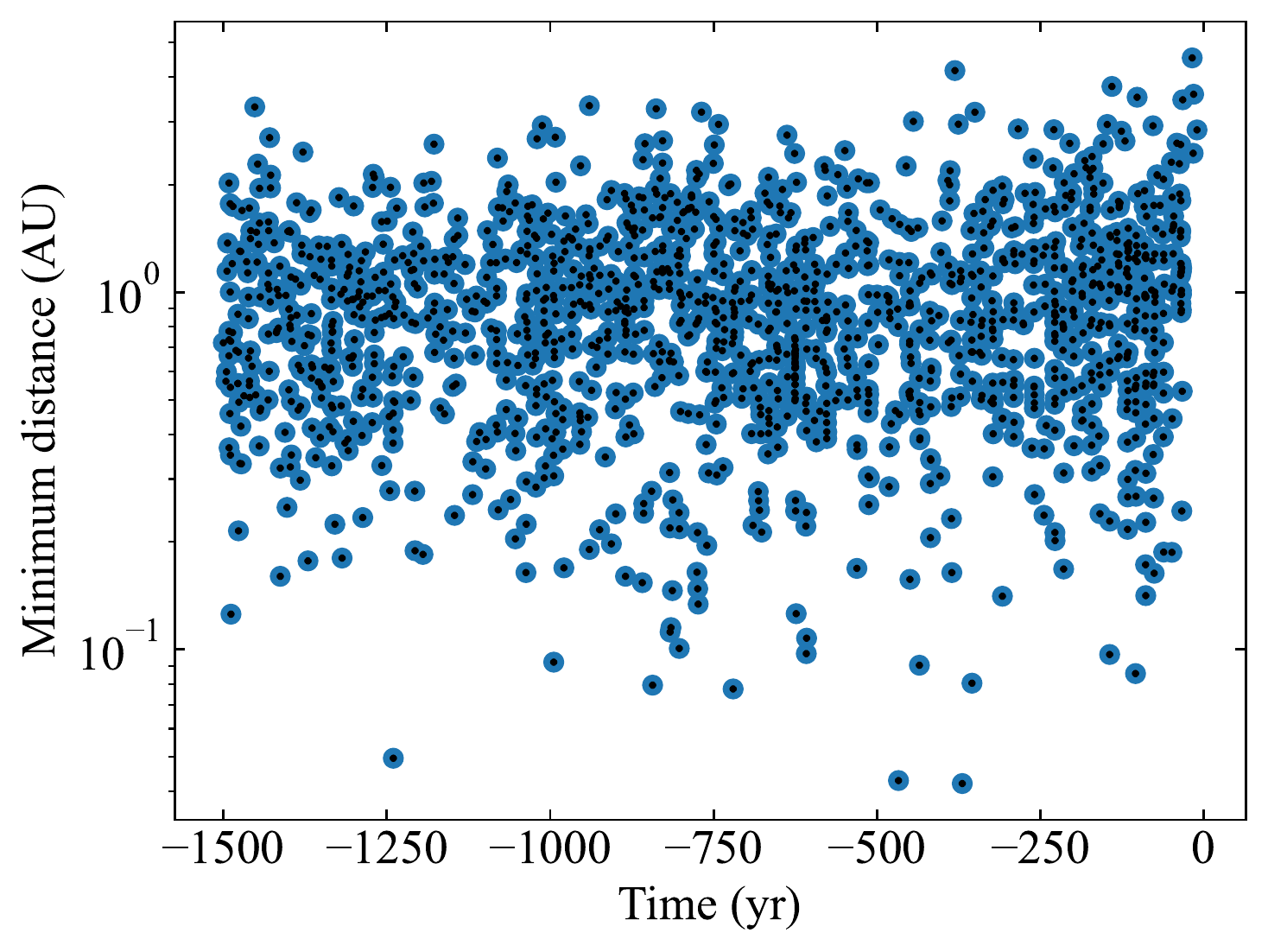}
        \includegraphics[width=0.49\linewidth]{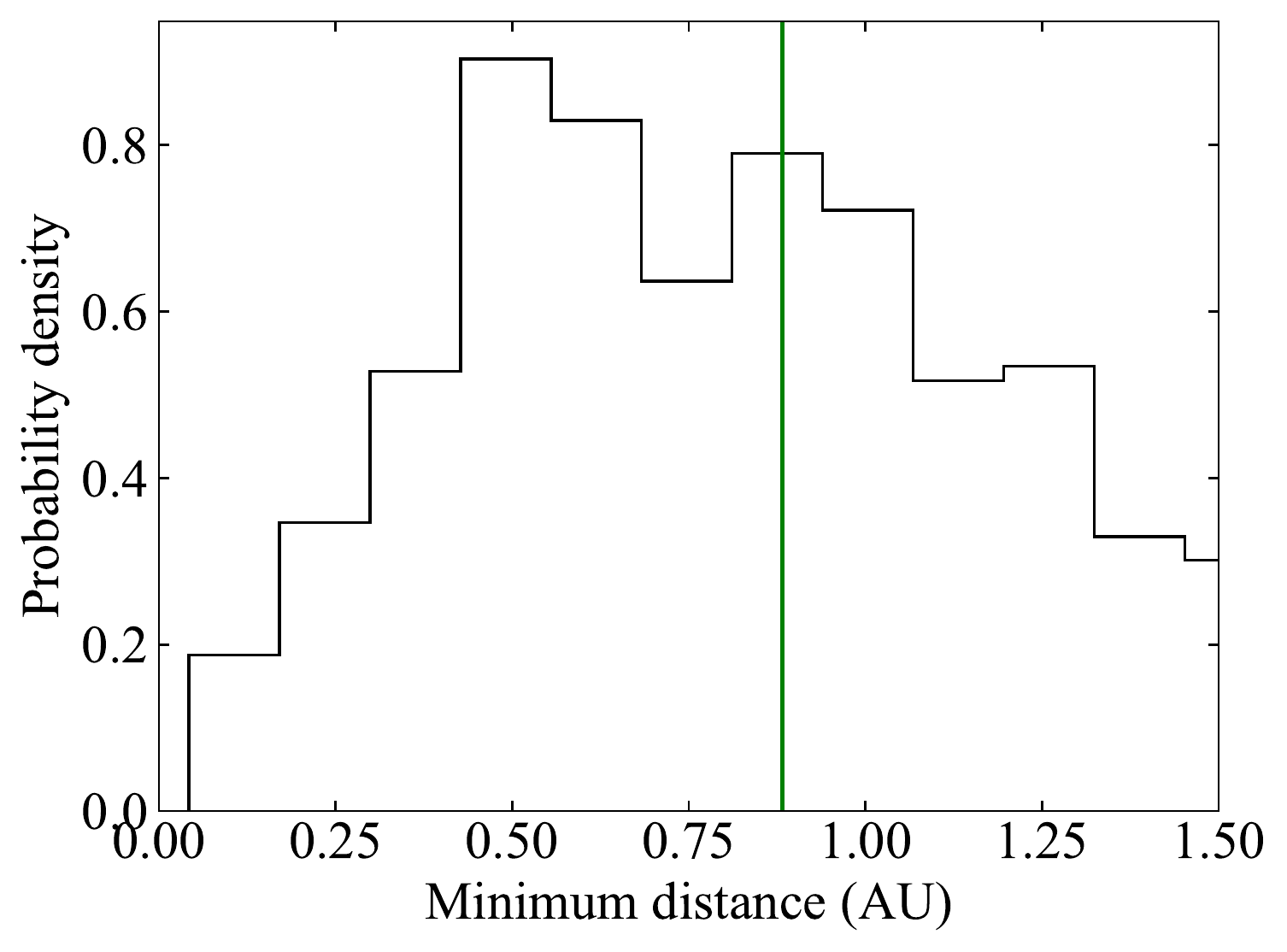}
        \includegraphics[width=0.49\linewidth]{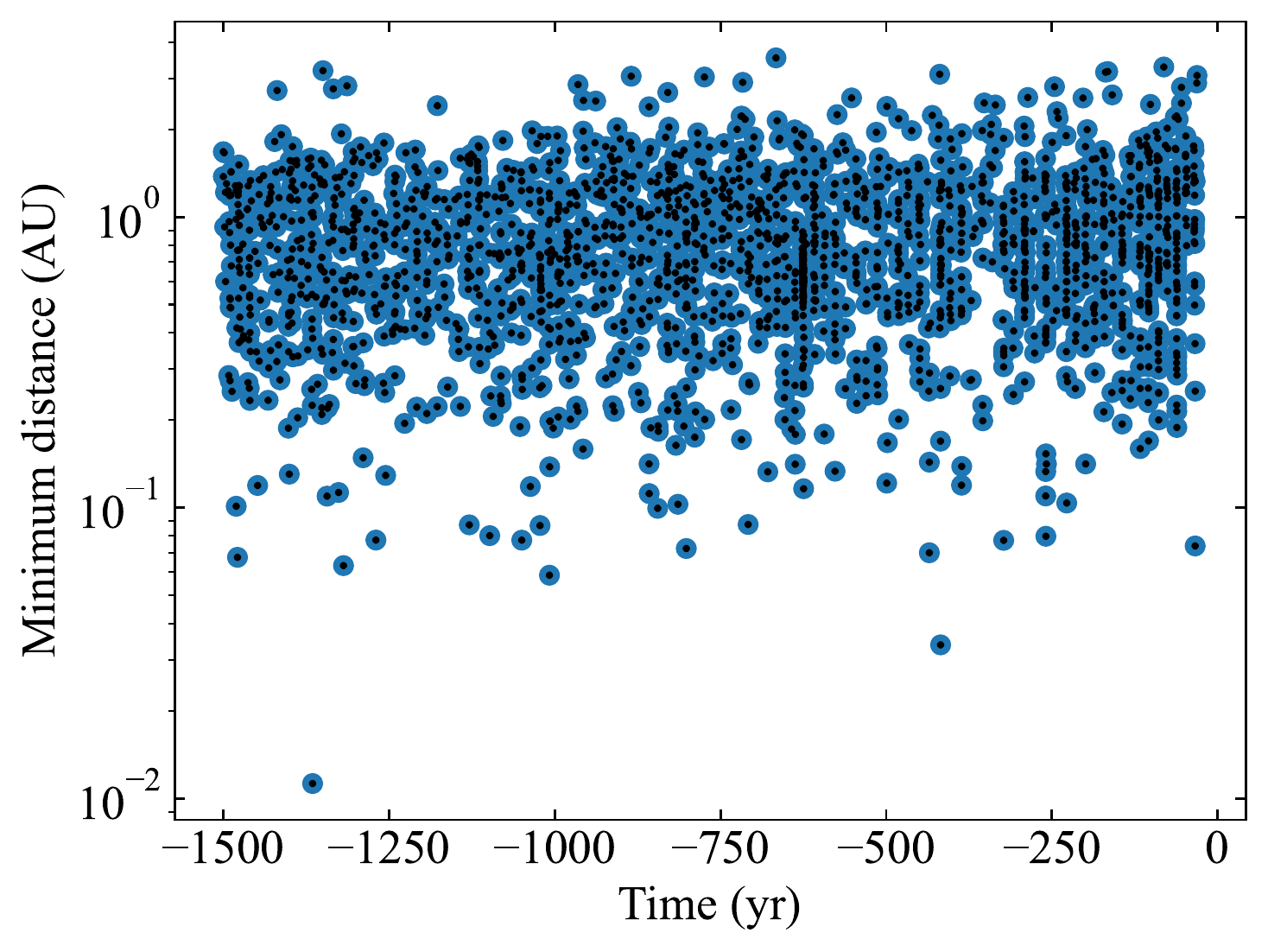}
        \includegraphics[width=0.49\linewidth]{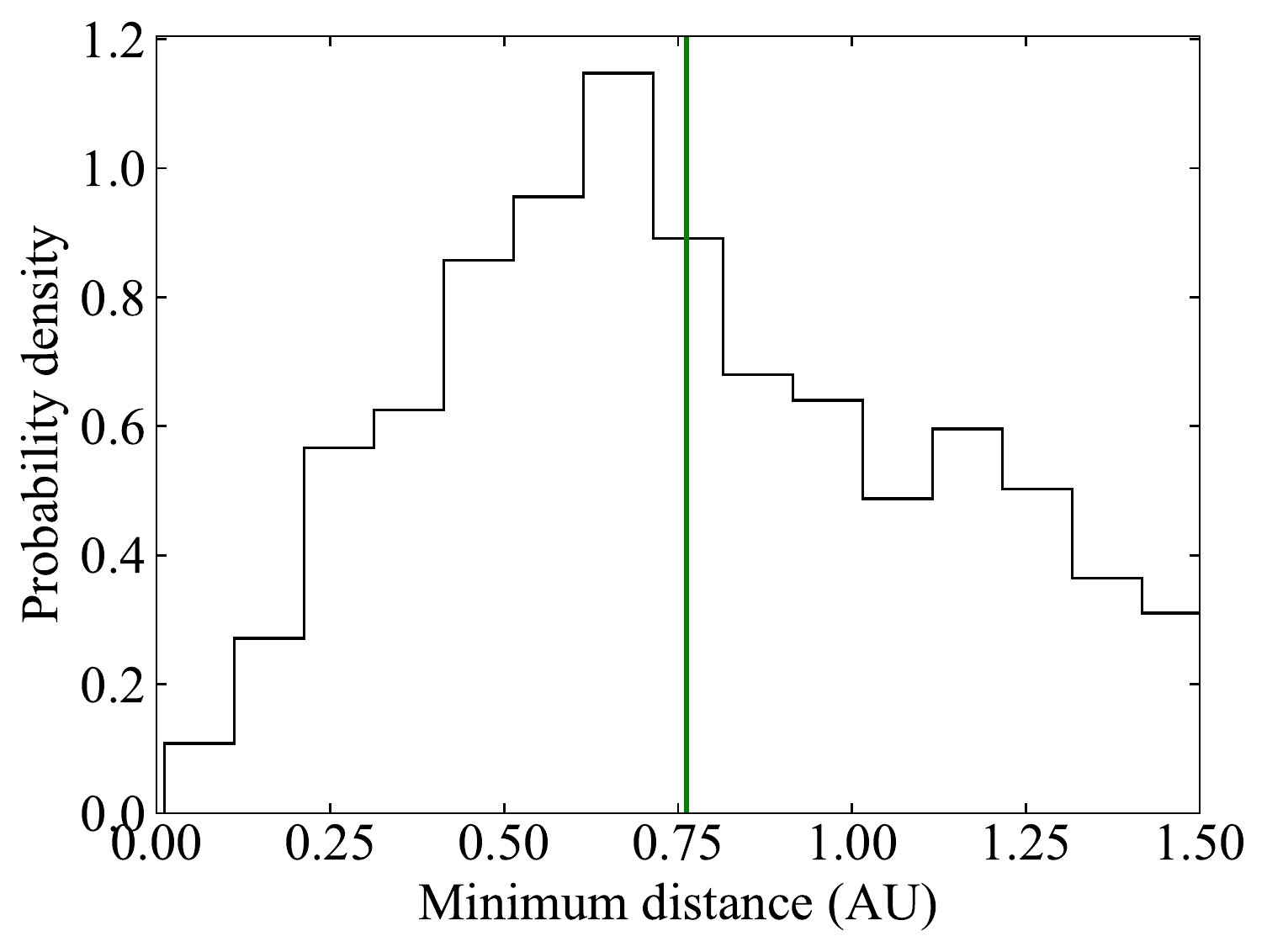}
        \caption{Distribution of minimum approach distances for the pair 2013~VZ$_{70}$ and Fornjot. {\em Top panels:} Assuming 
                 uncertainties of 10\% for the barycentric Cartesian state vector of Fornjot in Table~\ref{vectorFornjot}.  
                 {\em Bottom panels:} Assuming uncertainties of 5\%. The median values are shown as vertical green lines.
                }
        \label{Fornjot_encounters}
     \end{figure*}

\begin{figure*}
        \centering
        \includegraphics[width=0.49\linewidth]{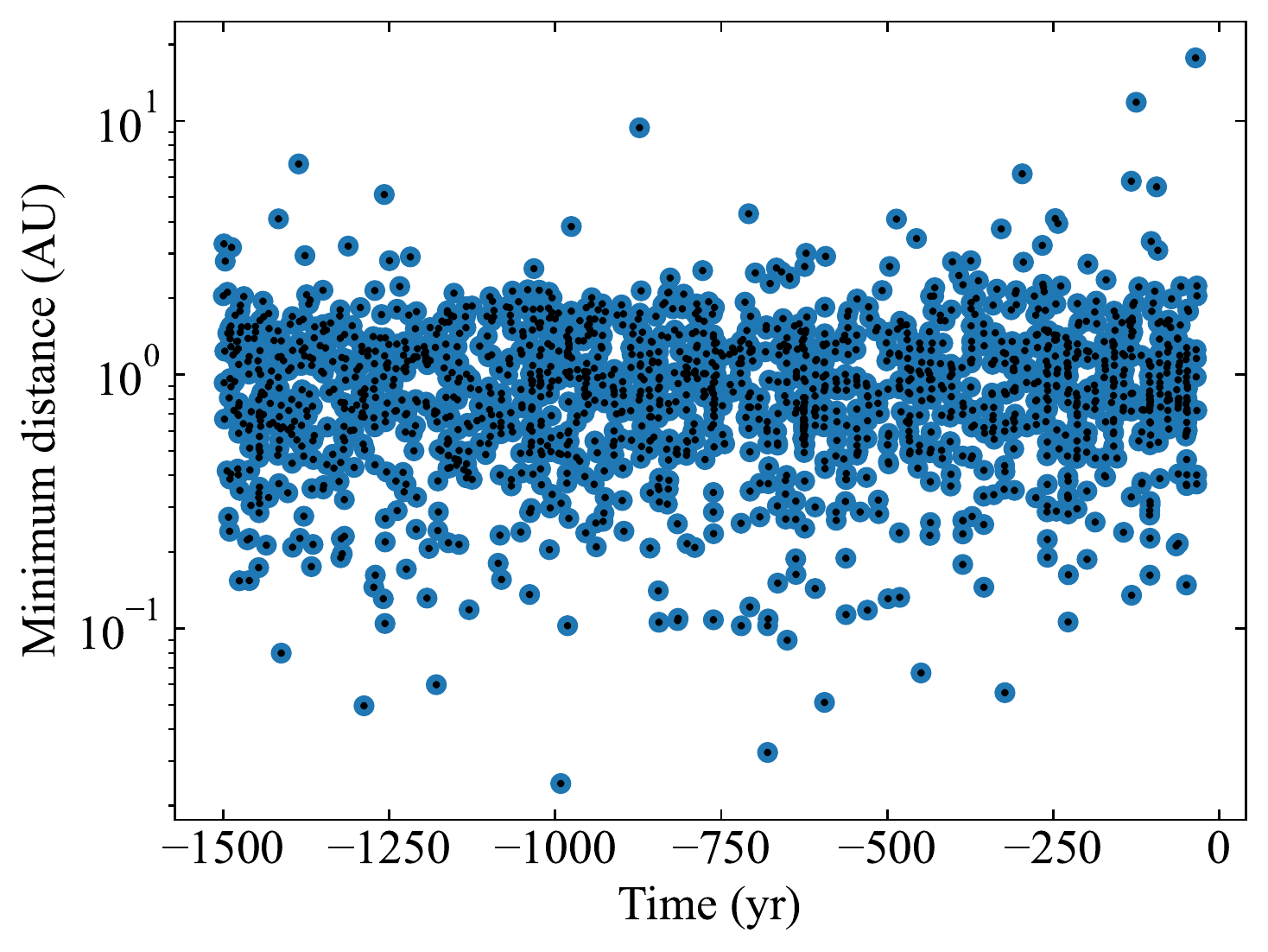}
        \includegraphics[width=0.49\linewidth]{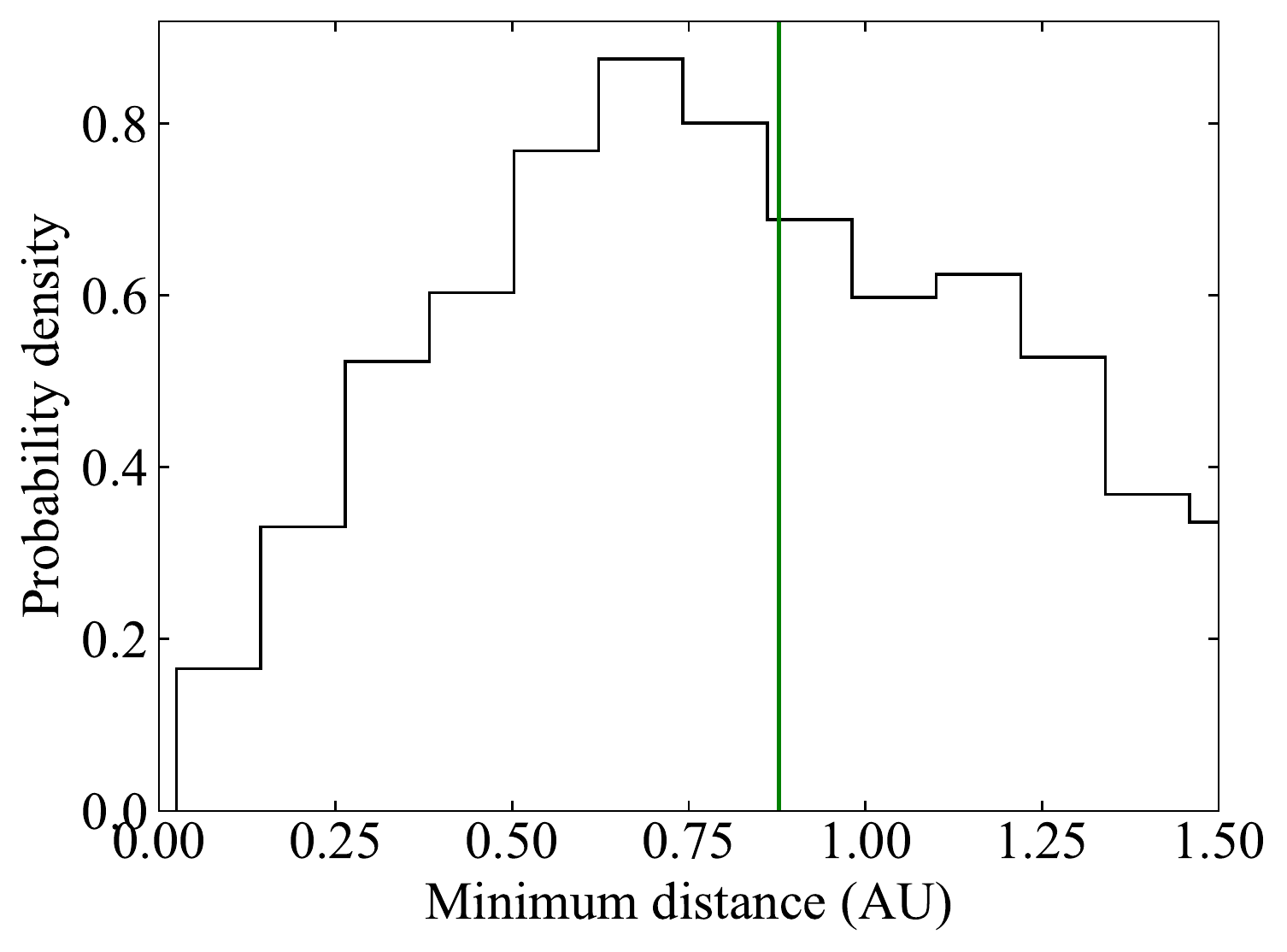}
        \includegraphics[width=0.49\linewidth]{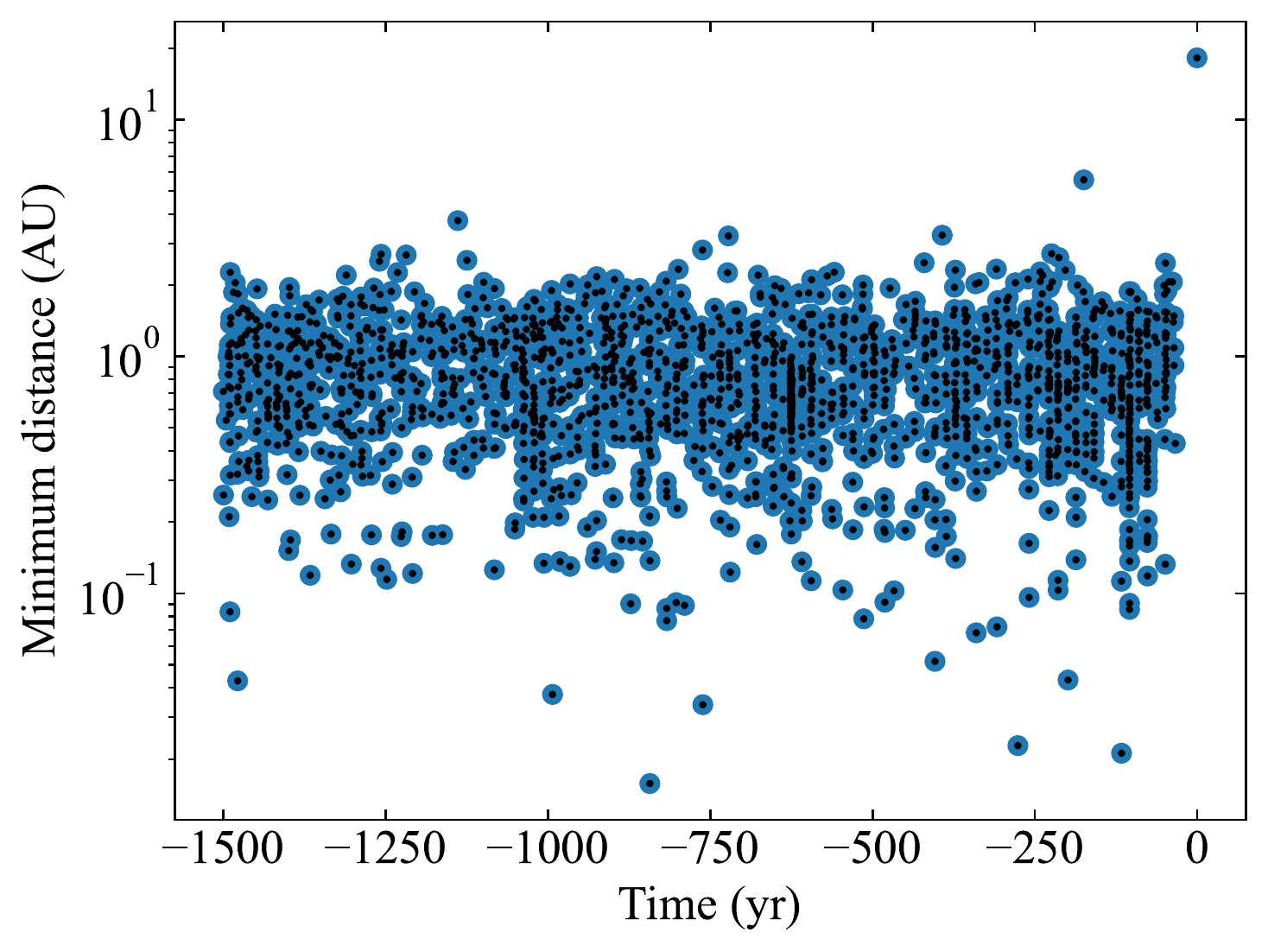}
        \includegraphics[width=0.49\linewidth]{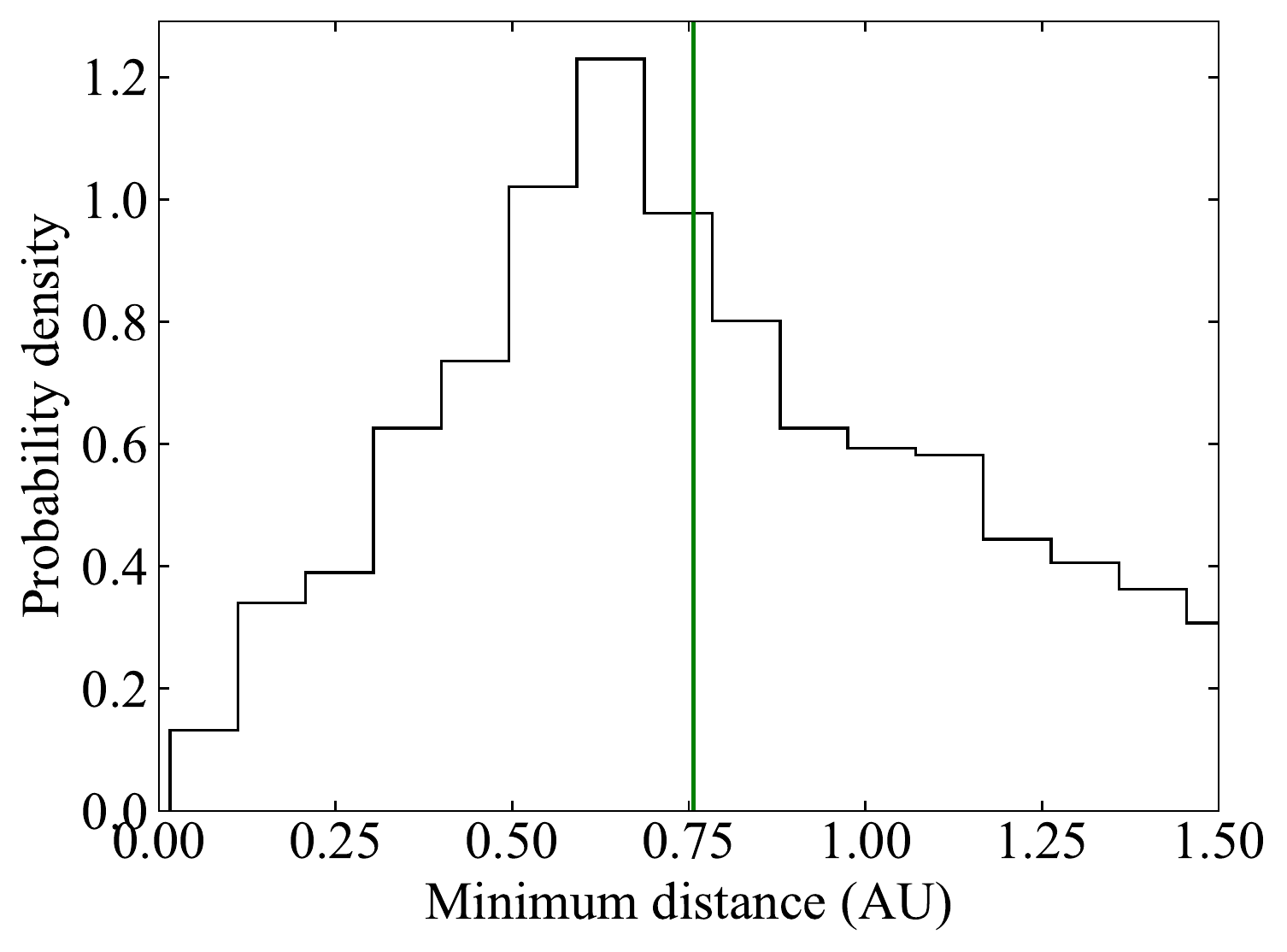}
        \caption{Distribution of minimum approach distances for the pair 2013~VZ$_{70}$ and Thrymr. {\em Top panels:} Assuming
                 uncertainties of 10\% for the barycentric Cartesian state vector of Thrymr in Table~\ref{vectorThrymr}.
                 {\em Bottom panels:} Assuming uncertainties of 5\%. The median values are shown as vertical green lines.
                }
        \label{Thrymr_encounters}
     \end{figure*}

         Figure~\ref{Fornjot_encounters} shows that flybys as close as 0.011~AU are possible, a result fully consistent with the one 
         obtained in the previous section for the distribution of mutual nodal distances. Our results also show that by considering smaller 
         uncertainties in the case of Fornjot, the minimum approach distance decreases. Figure~\ref{Thrymr_encounters} shows equivalent 
         results for Thrymr, and flybys as close as 0.016~AU are possible. These results, based on $N$-body calculations, confirm that 
         2013~VZ$_{70}$ may have approached Fornjot and Thrymr at relatively short range in the past, despite us using a rather conservative 
         choice when factoring uncertainties into the calculations. Lowering the level of uncertainty may produce flybys an order of 
         magnitude or two closer. In addition, our flyby experiments show that the brief periods of capture of 2013~VZ$_{70}$ as a 
         temporary irregular moon of Saturn are ubiquitous.
%
%
%
%
%
%
%
%

%
%
     \begin{figure*}
        \centering
        \includegraphics[width=0.49\linewidth]{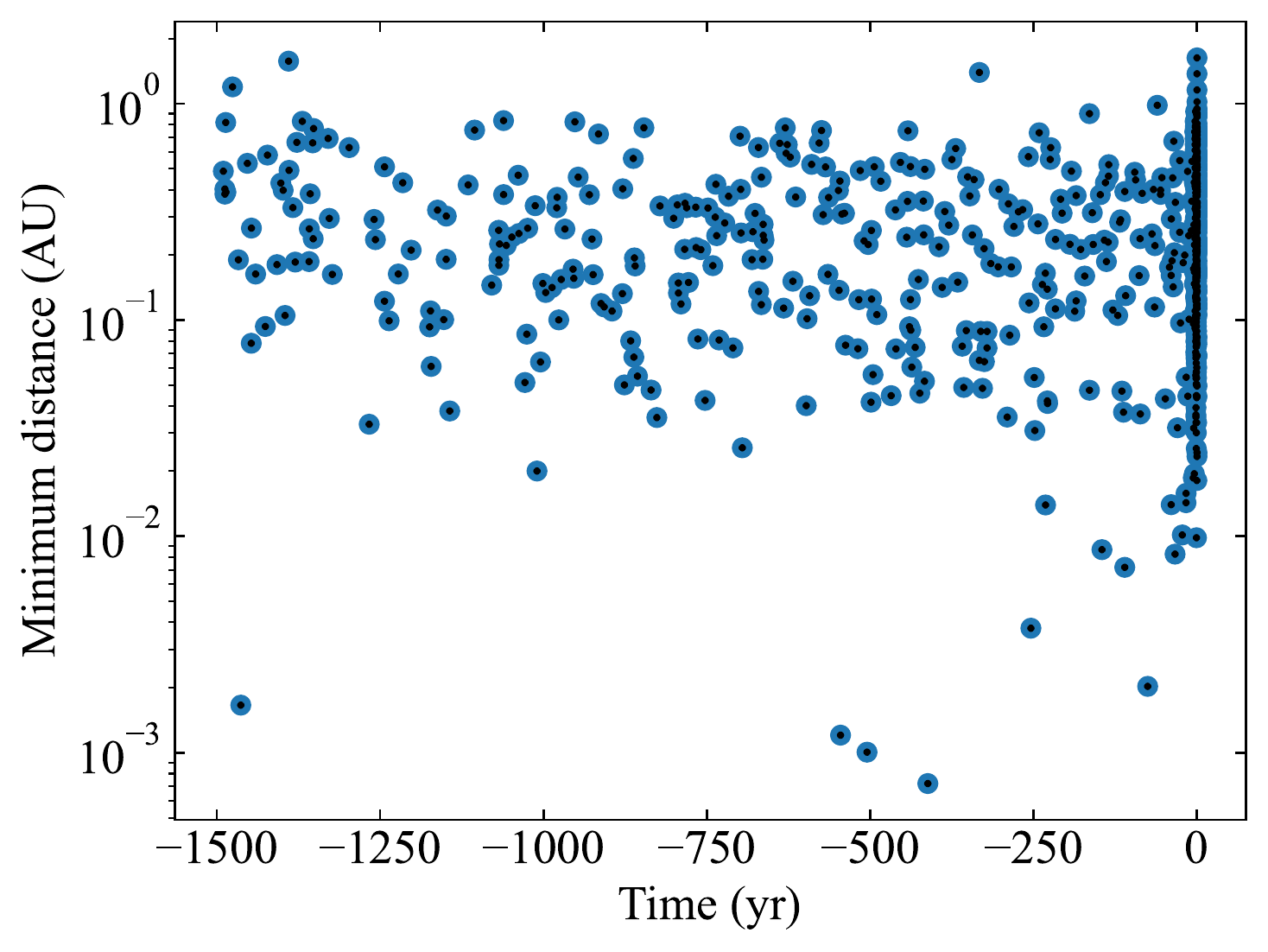}
        \includegraphics[width=0.49\linewidth]{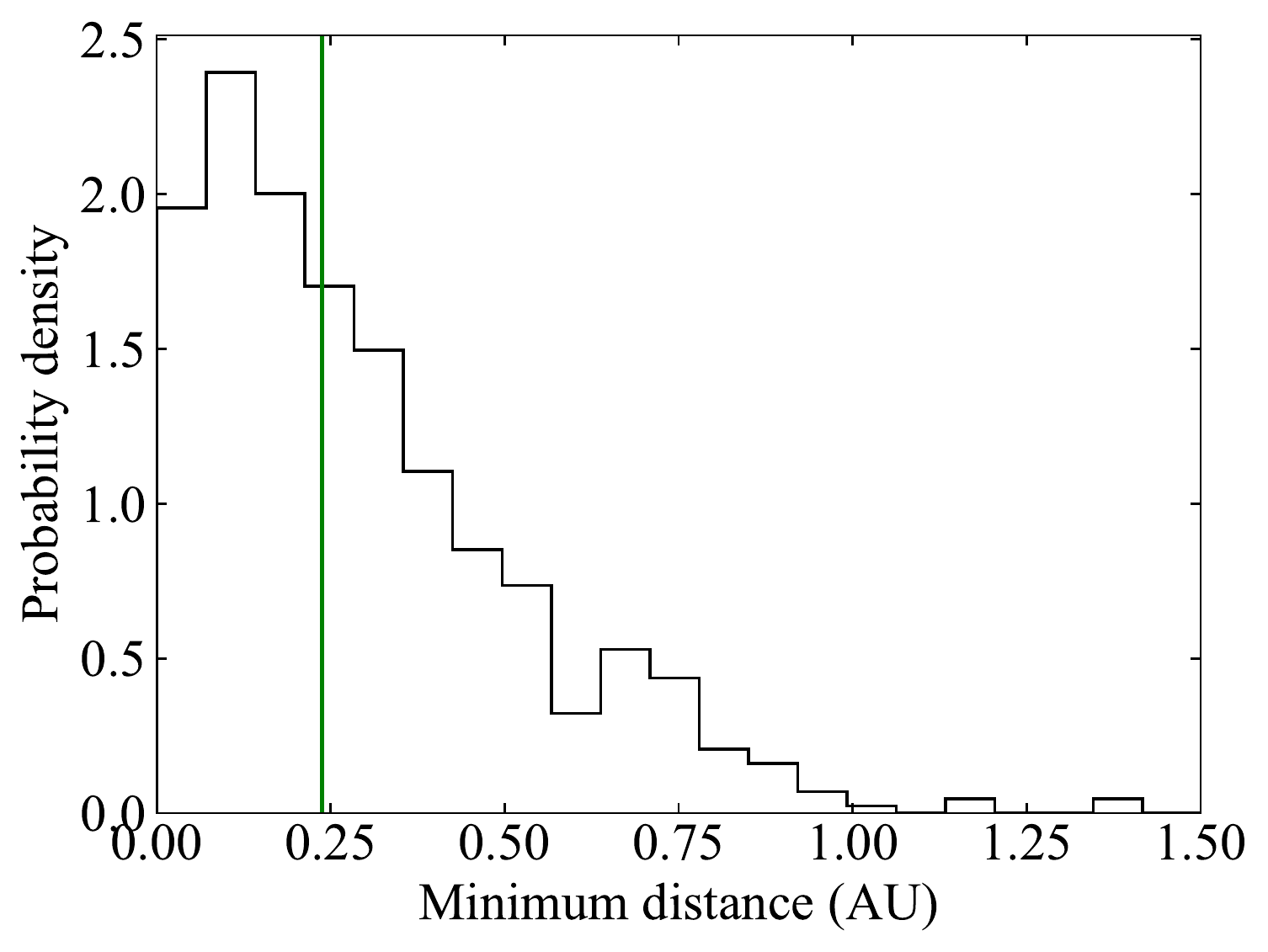}
        \caption{Distribution of minimum approach distances for the pair Fornjot and Thrymr. Our calculations assumed uncertainties of 5\%. 
                 The median values are shown as vertical green lines.
                }
        \label{moon_encounters}
     \end{figure*}
%
%
         As an additional quality control step, we repeated the experiment (600 instances), focusing on encounters between the moons 
         Fornjot and Thrymr. Figure~\ref{moon_encounters} shows that flybys as close as 0.0007~AU are possible when assuming uncertainties 
         of 5\%. 

   \section{Summary and conclusions\label{Conclusions}}
      In this work we have studied the orbital evolution of 2013~VZ$_{70}$ backward and forward in time using direct $N$-body simulations 
      and factoring the uncertainties into the calculations. We have also explored a possible connection between 2013~VZ$_{70}$ and the 
      moons of Saturn by computing the distribution of mutual nodal distances between this centaur and a sample of moons to investigate how 
      close two orbits can get to each other. Our conclusions can be summarized as follows.
      \begin{enumerate}
         \item We show that 2013~VZ$_{70}$ is a present-day co-orbital to Saturn of the horseshoe type. It is, however, a transient 
               co-orbital. 
         \item Centaur 2013~VZ$_{70}$ may approach Saturn at very low relative velocity; as a result, it might experience brief periods of 
               capture as a temporary irregular moon.
         \item The orbit of 2013~VZ$_{70}$ is similar in terms of eccentricity and inclination to the Inuit group of irregular 
               moons of Saturn, particularly Siarnaq and Tarqeq.
         \item The analysis of the distribution of mutual nodal distances between 2013~VZ$_{70}$ and a sample of moons shows that the 
               mutual nodal distance for the descending nodes of this centaur and Fornjot is 0.015~AU and that for the ascending nodes of
               the object and Thrymr is 0.058~AU. In both cases, the values are below the first percentile of the distribution. 
         \item The orbit determination of 2013~VZ$_{70}$ still requires significant improvement. Any prediction beyond the time interval 
               ($-$1000,~1000)~yr based on the orbit determination in Table~\ref{elements} is very uncertain. The available data cannot be
               used to confirm or reject an origin for 2013~VZ$_{70}$ in the trans-Neptunian populations or within the groups of irregular
               moons of Saturn, which have in some cases even more uncertain orbit determinations. 
      \end{enumerate}
      The facts are that (i) 2013~VZ$_{70}$ might pass fairly close (as confirmed with $N$-body calculations) to some of the moons of the 
      Norse group that could be debris from Phoebe \citep{2019Icar..322...80D}, (ii) its orbit is similar to the heliocentric orbits of some 
      of the moons of the Inuit group (see Tables~\ref{elements} and \ref{moonelements}), and (iii) it also can approach Saturn at low 
      relative velocity, sufficiently low to become a temporary moon itself, as discussed in Sect.~\ref{Results}. These three objective 
      pieces of information are supportive of a scenario of in situ formation via impact, fragmentation, or tidal disruption within the 
      population of the irregular moons of Saturn. The orbital evolution analysis in Sect.~\ref{Results} suggests that the putative 
      formation event may have taken place relatively recently or, because predictions are very uncertain, more than about 1000~yr ago. 

      On the other hand, if 2013~VZ$_{70}$ is debris linked to Phoebe, which is believed to be a captured object with an origin in 
      trans-Neptunian space (see for example \citealt{2005Natur.435...69J,2007ARA&A..45..261J} but also consider 
      \citealt{2019MNRAS.486..538C}), spectroscopic studies may not be able to confirm or refute its putative origin, captured versus 
      collisional. It is possible that only improvements in the orbit determinations of 2013~VZ$_{70}$ and the population of irregular moons 
      of Saturn will eventually lead to a robust solution of this dilemma.

   \begin{acknowledgements}
      We thank the referee for her/his prompt report that included a very helpful suggestion regarding the interpretation of our results, 
      S.~J. Aarseth for providing one of the codes used in this research, A.~B. Chamberlin for helping with the new JPL's Solar System 
      Dynamics website, J.~D. Giorgini for comments and insight on the uncertainties of the orbit determinations of the irregular moons of 
      Saturn, and A.~I. G\'omez de Castro for providing access to computing facilities. Part of the calculations and the data analysis were 
      completed on the Brigit HPC server of the `Universidad Complutense de Madrid' (UCM), and we thank S. Cano Als\'ua for his help during 
      this stage. This work was partially supported by the Spanish `Ministerio de Econom\'{\i}a y Competitividad' (MINECO) under grant 
      ESP2017-87813-R. In preparation of this paper, we made use of the NASA Astrophysics Data System, the ASTRO-PH e-print server, and the 
      MPC data server. 
   \end{acknowledgements}

   \bibliographystyle{aa}

\begin{thebibliography}{}
      \bibitem[Aarseth(2003)]{2003gnbs.book.....A} Aarseth, S.~J. 2003,
              Gravitational N-Body Simulations
              (Cambridge: Cambridge University Press), 27
      \bibitem[Alexandersen et al.(2018)]{2018DPS....5030509A} Alexandersen, M., Greenstreet, S., Gladman, B., et al.\ 2018, 
              in AAS/Division for Planetary Sciences Meeting Abstracts, Vol.~50, 305.09 
      \bibitem[Alexandersen et al.(2020)]{2020DPS....5220606A} Alexandersen, M., Greenstreet, S., Gladman, B., et al.\ 2020, 
              in AAS/Division for Planetary Sciences Meeting Abstracts, Vol.~52, 206.06 
      \bibitem[Alexandersen et al.(2021)]{2021PSJ.....2..212A} Alexandersen, M., Greenstreet, S., Gladman, B., et al.\ 2021, 
              PSJ, 2, 212 
      \bibitem[Ashton et al.(2021)]{2021PSJ.....2..158A} Ashton, E., Gladman, B., \& Beaudoin, M.\ 2021, 
              PSJ, 2, 158
      \bibitem[Bannister et al.(2016)]{2016AJ....152...70B} Bannister, M.~T., Kavelaars, J.~J., Petit, J.-M., et al.\ 2016, 
              \aj, 152, 70
      \bibitem[Bannister et al.(2021)]{2021MPEC....Q...55B} Bannister, M.~T., Kavelaars, J.~J., Gladman, B.~J., et al.\ 2021, 
              Minor Planet Electronic Circulars, 2021-Q55
      \bibitem[Castillo-Rogez et al.(2019)]{2019MNRAS.486..538C} Castillo-Rogez, J., Vernazza, P., \& Walsh, K.\ 2019, 
              \mnras, 486, 538
      \bibitem[de la Barre et al.(1996)]{1996Icar..121...88D} de la Barre, C.~M., Kaula, W.~M., \& Varadi, F.\ 1996, 
              \icarus, 121, 88
      \bibitem[de la Fuente Marcos \& de la Fuente Marcos(2012)]{2012MNRAS.427..728D} de la Fuente Marcos, C. \& de la Fuente Marcos, R.\ 2012, 
              \mnras, 427, 728
      \bibitem[de la Fuente Marcos \& de la Fuente Marcos(2016)]{2016MNRAS.462.3344D} de la Fuente Marcos, C. \& de la Fuente Marcos, R.\ 2016, 
              \mnras, 462, 3344
      \bibitem[de la Fuente Marcos \& de la Fuente Marcos(2021a)]{2021MNRAS.501.6007D} de la Fuente Marcos, C. \& de la Fuente Marcos, R.\ 2021a, 
              \mnras, 501, 6007
      \bibitem[de la Fuente Marcos et al.(2021b)]{2021A&A...649A..85D} de la Fuente Marcos, C., de la Fuente Marcos, R., Licandro, J., et al.\ 2021b, 
              \aap, 649, A85
      \bibitem[Denk \& Mottola(2019)]{2019Icar..322...80D} Denk, T. \& Mottola, S.\ 2019, 
              \icarus, 322, 80
      \bibitem[Dermott \& Murray(1981)]{1981Icar...48...12D} Dermott, S.~F. \& Murray, C.~D.\ 1981, 
              \icarus, 48, 12
      \bibitem[Everhart(1973)]{1973AJ.....78..316E} Everhart, E.\ 1973, 
              \aj, 78, 316 
      \bibitem[Fedorets et al.(2017)]{2017Icar..285...83F} Fedorets, G., Granvik, M., \& Jedicke, R.\ 2017, 
              \icarus, 285, 83
      \bibitem[{Freedman \& Diaconis(1981)}]{Freedman1981} Freedman, D. \& Diaconis, P. 1981, 
              Zeitschrift f{\"u}r Wahrscheinlichkeitstheorie und Verwandte Gebiete, 57, 453
      \bibitem[Gallardo(2006)]{2006Icar..184...29G} Gallardo, T.\ 2006, 
              \icarus, 184, 29
      \bibitem[Ginsburg et al.(2019)]{2019AJ....157...98G} Ginsburg, A., Sip{\H{o}}cz, B.~M., Brasseur, C.~E., et al.\ 2019,
              \aj, 157, 98
      \bibitem[{{Giorgini}(2011)}]{2011jsrs.conf...87G} {Giorgini}, J. 2011, 
              in Journ\'ees Syst\`emes de R\'ef\'erence Spatio-temporels 2010, 
              ed. N.~{Capitaine}, 87--87 
      \bibitem[Giorgini(2015)]{2015IAUGA..2256293G} Giorgini, J.~D.\ 2015,
              IAUGA, 22, 2256293
      \bibitem[Gladman et al.(2000)]{2000IAUC.7513....1G} Gladman, B., Kavelaars, J., Allen, R.~L., et al.\ 2000, 
              \iaucirc, 7513
      \bibitem[Grav et al.(2015)]{2015ApJ...809....3G} Grav, T., Bauer, J.~M., Mainzer, A.~K., et al.\ 2015, 
              \apj, 809, 3
      \bibitem[Harris et al.(2020)]{2020Natur.585..357H} Harris, C.~R., Millman, K.~J., van der Walt, S.~J., et al.\ 2020, 
              \nat, 585, 357
      \bibitem[Holman et al.(2001)]{2001MPEC....U...42H} Holman, M., Gladman, B., Grav, T., et al.\ 2001, 
              Minor Planet Electronic Circulars, 2001-U42
      \bibitem[Hou et al.(2014)]{2014MNRAS.437.1420H} Hou, X.~Y., Scheeres, D.~J., \& Liu, L.\ 2014, 
              \mnras, 437, 1420
      \bibitem[Huang et al.(2019)]{2019MNRAS.488.2543H} Huang, Y., Li, M., Li, J., et al.\ 2019, 
              \mnras, 488, 2543
      \bibitem[Hunter(2007)]{2007CSE.....9...90H} Hunter, J.~D.\ 2007,
              Computing in Science and Engineering, 9, 90 
      \bibitem[Innanen \& Mikkola(1989)]{1989AJ.....97..900I} Innanen, K.~A. \& Mikkola, S.\ 1989, 
              \aj, 97, 900
      \bibitem[Ito \& Tanikawa(1999)]{1999Icar..139..336I} Ito, T. \& Tanikawa, K.\ 1999, 
              \icarus, 139, 336 
      \bibitem[Ito \& Tanikawa(2002)]{2002MNRAS.336..483I} Ito, T. \& Tanikawa, K.\ 2002, 
              \mnras, 336, 483
      \bibitem[Jacobson et al.(2012)]{2012AJ....144..132J} Jacobson, R., Brozovi{\'c}, M., Gladman, B., et al.\ 2012, 
              \aj, 144, 132
      \bibitem[Jedicke et al.(2018)]{2018FrASS...5...13J} Jedicke, R., Bolin, B.~T., Bottke, W.~F., et al.\ 2018, 
              Frontiers in Astronomy and Space Sciences, 5, 13
      \bibitem[Jewitt \& Haghighipour(2007)]{2007ARA&A..45..261J} Jewitt, D. \& Haghighipour, N.\ 2007, 
              \araa, 45, 261
      \bibitem[Jewitt et al.(2006)]{2006MPEC....C...74J} Jewitt, D.~C., Sheppard, S.~S., Kleyna, J., et al.\ 2006, 
              Minor Planet Electronic Circulars, 2006-C74
      \bibitem[Johnson \& Lunine(2005)]{2005Natur.435...69J} Johnson, T.~V. \& Lunine, J.~I.\ 2005, 
              \nat, 435, 69
      \bibitem[Li et al.(2018)]{2018A&A...617A.114L} Li, M., Huang, Y., \& Gong, S.\ 2018, 
              \aap, 617, A114
      \bibitem[Makino(1991)]{1991ApJ...369..200M} Makino, J.\ 1991,
              \apj, 369, 200
      \bibitem[Marzari et al.(2002)]{2002ApJ...579..905M} Marzari, F., Tricarico, P., \& Scholl, H.\ 2002, 
              \apj, 579, 905
      \bibitem[Melita \& Brunini(2001)]{2001MNRAS.322L..17M} Melita, M.~D. \& Brunini, A.\ 2001, 
              \mnras, 322, L17 
      \bibitem[Mikkola et al.(2006)]{2006MNRAS.369...15M} Mikkola, S., Innanen, K., Wiegert, P., et al.\ 2006, 
              \mnras, 369, 15 
      \bibitem[Milani et al.(1989)]{1989Icar...82..200M} Milani, A., Nobili, A.~M., \& Carpino, M.\ 1989,
              \icarus, 82, 200
      \bibitem[Morais \& Namouni(2013a)]{2013CeMDA.117..405M} Morais, M.~H.~M. \& Namouni, F.\ 2013a, 
              Celestial Mechanics and Dynamical Astronomy, 117, 405 
      \bibitem[Morais \& Namouni(2013b)]{2013MNRAS.436L..30M} Morais, M.~H.~M. \& Namouni, F.\ 2013b, 
              \mnras, 436, L30
      \bibitem[Morais \& Namouni(2017)]{2017Natur.543..635M} Morais, H. \& Namouni, F.\ 2017, 
              \nat, 543, 635
      \bibitem[Morais \& Namouni(2019)]{2019MNRAS.490.3799M} Morais, M.~H.~M. \& Namouni, F.\ 2019, 
              \mnras, 490, 3799
      \bibitem[Murray \& Dermott(1999)]{1999ssd..book.....M} Murray, C.~D., \& Dermott, S.~F.\ 1999,
              Solar System Dynamics
              (Cambridge: Cambridge University Press)
      \bibitem[Musen(1971)]{1971NASTN6279.....M} Musen, P.\ 1971, 
              NASA Tech. Note, 6279
      \bibitem[Namouni(1999)]{1999Icar..137..293N} Namouni, F.\ 1999, 
              \icarus, 137, 293
      \bibitem[Namouni et al.(1999)]{1999PhRvL..83.2506N} Namouni, F., Christou, A.~A., \& Murray, C.~D.\ 1999, 
              \prl, 83, 2506
      \bibitem[Namouni \& Murray(2000)]{2000CeMDA..76..131N} Namouni, F. \& Murray, C.~D.\ 2000, 
              Celestial Mechanics and Dynamical Astronomy, 76, 131
      \bibitem[Nesvorn{\'y} \& Dones(2002)]{2002Icar..160..271N} Nesvorn{\'y}, D. \& Dones, L.\ 2002, 
              \icarus, 160, 271
      \bibitem[Pedregosa et al.(2011)]{2012arXiv1201.0490P} Pedregosa, F., Varoquaux, G., Gramfort, A., et al.\ 2011,
              Journal of Machine Learning Research, 12, 2825
      \bibitem[Petit et al.(2001)]{2001MPEC....X...20P} Petit, J.-M., Nicholson, P., Dumas, C., et al.\ 2001, 
              Minor Planet Electronic Circulars, 2001-X20
      \bibitem[Saillenfest et al.(2017)]{2017CeMDA.129..329S} Saillenfest, M., Fouchard, M., Tommei, G., et al.\ 2017, 
              Celestial Mechanics and Dynamical Astronomy, 129, 329
      \bibitem[Sheppard et al.(2007a)]{2007IAUC.8836....1S} Sheppard, S.~S., Jewitt, D.~C., Kleyna, J., et al.\ 2007a, 
              \iaucirc, 8836
      \bibitem[Sheppard et al.(2007b)]{2007MPEC....G...38S} Sheppard, S.~S., Jewitt, D.~C., Kleyna, J., et al.\ 2007b, 
              Minor Planet Electronic Circulars, 2007-G38
      \bibitem[Sidorenko(2020)]{2020AJ....160..257S} Sidorenko, V.~V.\ 2020, 
              \aj, 160, 257
      \bibitem[Sidorenko et al.(2014)]{2014CeMDA.120..131S} Sidorenko, V.~V., Neishtadt, A.~I., Artemyev, A.~V., et al.\ 2014, 
              Celestial Mechanics and Dynamical Astronomy, 120, 131
      \bibitem[Tanikawa \& Ito(2007)]{2007PASJ...59..989T} Tanikawa, K. \& Ito, T.\ 2007, 
              \pasj, 59, 989
      \bibitem[Turrini et al.(2008)]{2008MNRAS.391.1029T} Turrini, D., Marzari, F., \& Beust, H.\ 2008, 
              \mnras, 391, 1029
      \bibitem[Turrini et al.(2009)]{2009MNRAS.392..455T} Turrini, D., Marzari, F., \& Tosi, F.\ 2009, 
              \mnras, 392, 455 
      \bibitem[van der Walt et al.(2011)]{2011CSE....13b..22V} van der Walt, S., Colbert, S.~C., \& Varoquaux, G.\ 2011, 
              Computing in Science and Engineering, 13, 22
      \bibitem[Wan et al.(2001)]{2001AJ....121.1155W} Wan, X.-S., Huang, T.-Y., \& Innanen, K.~A.\ 2001,
              \aj, 121, 1155
      \bibitem[Wiegert et al.(2000)]{2000AJ....119.1978W} Wiegert, P., Innanen, K., \& Mikkola, S.\ 2000, 
              \aj, 119, 1978
      \bibitem[Zink et al.(2020)]{2020AJ....160..232Z} Zink, J.~K., Batygin, K., \& Adams, F.~C.\ 2020, 
              \aj, 160, 232
   \end{thebibliography}

   \begin{appendix}

      \section{Input data\label{Adata}}
         Here, we include the barycentric Cartesian state vectors of the centaur 2013~VZ$_{70}$ and the moons Fornjot and Thrymr.
         These vectors and their uncertainties (provided or assumed) were used to carry out the calculations discussed in the main text, to generate the figures that display the time evolution of the critical angle, and to generate the histograms and distributions 
         of the close encounters of pairs of objects. For example, a new value of the $X$ component of the state vector is 
         computed as $X_{\rm c} = X + \sigma_X \ r$, where $r$ is a univariate Gaussian random number and $X$ and $\sigma_X$ are 
         the mean value and its 1$\sigma$ uncertainty (provided or assumed) in the corresponding table.
%
%
     \begin{table}
      \centering
      \fontsize{8}{12pt}\selectfont
      \tabcolsep 0.15truecm
      \caption{\label{vector2013VZ70}Barycentric Cartesian state vector of 2013~VZ$_{70}$: components and associated 1$\sigma$
               uncertainties.
              }
      \begin{tabular}{ccc}
       \hline
        Component                         &   &    value$\pm$1$\sigma$ uncertainty                                \\
       \hline
        $X$ (AU)                          & = & $-$6.511583173348614$\times10^{+0}$$\pm$1.35733831$\times10^{-4}$ \\
        $Y$ (AU)                          & = &    5.141776391203631$\times10^{+0}$$\pm$6.09013204$\times10^{-4}$ \\
        $Z$ (AU)                          & = & $-$1.696420134894827$\times10^{+0}$$\pm$9.77589520$\times10^{-5}$ \\
        $V_X$ (AU/d)                      & = & $-$4.121981142209526$\times10^{-3}$$\pm$2.18437943$\times10^{-7}$ \\
        $V_Y$ (AU/d)                      & = & $-$4.543278515226239$\times10^{-3}$$\pm$4.04638717$\times10^{-7}$ \\
        $V_Z$ (AU/d)                      & = &    2.854425621190691$\times10^{-4}$$\pm$9.65108764$\times10^{-8}$ \\
       \hline
      \end{tabular}
      \tablefoot{Data are referred to epoch JD 2459396.5, which corresponds to 0:00 on 2021 July 1 TDB  (J2000.0 ecliptic and 
                 equinox). Source: JPL's {\tt Horizons}. 
                }
     \end{table}
%
%
%
%
     \begin{table}
      \centering
      \fontsize{8}{12pt}\selectfont
      \tabcolsep 0.15truecm
      \caption{\label{vectorFornjot}Barycentric Cartesian state vector of Fornjot (originally named S/2004 S 8 or Saturn XLII): 
               components.
              }
      \begin{tabular}{ccc}
       \hline
        Component                         &   &    value                            \\
       \hline
        $X$ (AU)                          & = &    6.398591117555700$\times10^{+0}$ \\
        $Y$ (AU)                          & = & $-$7.714490185713658$\times10^{+0}$ \\
        $Z$ (AU)                          & = & $-$1.478426583748547$\times10^{-1}$ \\
        $V_X$ (AU/d)                      & = &    4.029956027408442$\times10^{-3}$ \\
        $V_Y$ (AU/d)                      & = &    2.762140815948923$\times10^{-3}$ \\
        $V_Z$ (AU/d)                      & = & $-$1.756170017563706$\times10^{-4}$ \\
       \hline
      \end{tabular}
      \tablefoot{Data are referred to epoch JD 2459396.5, which corresponds to 0:00 on 2021 July 1 TDB  (J2000.0 ecliptic and 
                 equinox). Source: JPL's {\tt Horizons}. 
                }
     \end{table}
%
%
%
%
     \begin{table}
      \centering
      \fontsize{8}{12pt}\selectfont
      \tabcolsep 0.15truecm
      \caption{\label{vectorThrymr}Barycentric Cartesian state vector of Thrymr (originally named S/2000 S 7 or Saturn XXX): 
               components.
              }
      \begin{tabular}{ccc}
       \hline
        Component                         &   &    value                            \\
       \hline
        $X$ (AU)                          & = &    6.344914469366030$\times10^{+0}$ \\
        $Y$ (AU)                          & = & $-$7.652929926514768$\times10^{+0}$ \\
        $Z$ (AU)                          & = & $-$1.241275465603603$\times10^{-1}$ \\
        $V_X$ (AU/d)                      & = &    4.742634945291553$\times10^{-3}$ \\
        $V_Y$ (AU/d)                      & = &    3.186493033277793$\times10^{-3}$ \\
        $V_Z$ (AU/d)                      & = & $-$2.948191331774928$\times10^{-4}$ \\
       \hline
      \end{tabular}
      \tablefoot{Data are referred to epoch JD 2459396.5, which corresponds to 0:00 on 2021 July 1 TDB  (J2000.0 ecliptic and 
                 equinox). Source: JPL's {\tt Horizons}. 
                }
     \end{table}
%
%

   \end{appendix}

\end{document}